\documentclass[preprint, pteplogo]{ptephy_v2}

\preprintnumber{YITP-26-50, RIKEN-iTHEMS-Report-26}

\usepackage{amsmath, amssymb, amsthm}
\usepackage{mathrsfs}
\usepackage{physics}
\usepackage{graphicx}
\usepackage[dvipsnames]{xcolor}
\usepackage{tikz}
\usepackage{bbold, bm, bbm}
\usepackage{hyperref, url}
\usepackage[english]{babel}
\usepackage{alltt}
\usepackage{booktabs, multirow, quotes}
\usepackage{framed}
\usepackage[normalem]{ulem}

\newcommand{\beq}{\begin{equation}}
\newcommand{\eeq}{\end{equation}}
\newcommand{\bea}{\begin{eqnarray}}
\newcommand{\ea}{\end{eqnarray}}

\theoremstyle{remark}
\newtheorem{remark}{Remark}[section]

\begin{document}

\title{A Monte Carlo Study of the Dipolar Universality Class in Three Dimensions}

\author{Akira Matsumoto}
\affil{Graduate School of Science, Osaka Metropolitan University, 3-3-138 Sugimoto, Sumiyoshi-ku, Osaka 558-8585, Japan 
\email{akira.matsumoto@omu.ac.jp}}
\affil{RIKEN Center for Interdisciplinary Theoretical and Mathematical Sciences (iTHEMS), RIKEN, 2-1 Hirosawa, Wako, Saitama 351-0198, Japan}

\author{Yu Nakayama}
\author{Toshiki Onagi}
\affil{Yukawa Institute for Theoretical Physics, Kyoto University, Kitashirakawa Oiwakecho, Sakyo-ku, Kyoto 606-8502 Japan}

\author{Slava Rychkov}
\affil{Institut des Hautes \'Etudes Scientifiques, 91440 Bures-sur-Yvette, France}

\begin{abstract}
The dipolar universality class describes the phase transition in 3D ferromagnets with strong dipolar interactions, as first discussed by Aharony and Fisher in the 1970s.
While this universality class has been studied theoretically using renormalization group methods, as well as experimentally, little is known about it from Monte Carlo simulations.
In this paper, we aim to bridge this gap. We introduce a lattice model that faithfully implements the transverse constraint on the order parameter.
We introduce a Markov Chain Monte Carlo algorithm which involves a combination of local Metropolis updates preserving the constraint, and a global update of the zero mode.
We perform simulations on cubic lattices up to volume $48\times 48 \times 48$.
We observe a continuous phase transition between the disordered and ordered phases.
We obtain estimates of the critical exponent $\nu$ and the critical Binder ratio, and compare them with results from other techniques.
We also study whether rotational invariance emerges at the critical point, and discuss the influence of the toroidal geometry on this question.
\end{abstract}

\subjectindex{B38, I70}

\maketitle

\section{Introduction}

The Heisenberg universality class of the thermodynamic phase transitions in 3D ferromagnets with $O(3)$ symmetry has been widely studied theoretically 
\cite{Wilson:1973jj, Pelissetto:2000ek, Campostrini:2002ky, Kos:2016ysd, Hasenbusch:2020pwj, Chester:2020iyt, Hasenbusch:2023fmn} 
as well as experimentally (see \cite{Collins1989, Pelissetto:2000ek} and references therein).
The influence of perturbing interactions on these phase transitions is of significant interest 
\cite{aharony1976dependence, Collins1989, Gimenez-Grau:2023lpz}.

One such effect is the dipolar interaction.
It is inevitably present in real-world ferromagnets because we live in a world with a dynamical electromagnetic field.
Thus, a magnetic moment at a point sources a magnetic field which then acts on faraway magnetic moments, with intensity decreasing as a power law.
That the Heisenberg model does not include this term, but only the nearest-neighbor interaction, is an oversimplification.

Aharony and Fisher \cite{FisherAharony, AharonyFisher, AF-PRBII} pointed out that once the dipolar interaction term is added, it causes a crossover to a new "dipolar" universality class.
Its order parameter transforms as a vector under the spatial rotation symmetry, which can be thought of as the diagonal subgroup of the product of the global and spatial symmetries of the Heisenberg fixed point.
Furthermore, the order parameter is divergence-free.
These unusual features are responsible for the lack of conformal invariance of the model \cite{Gimenez-Grau:2023lpz}.
This precludes the use of conformal bootstrap methods \cite{Poland:2018epd}, and most theoretical studies of the dipolar universality class use the renormalization group, either perturbative \cite{PhysRevB.10.2078, 2022NuPhB.98515990K} or non-perturbative \cite{Nakayama:2023wrx, Kalagov_2026}.
This phase transition has also been studied experimentally, using ferromagnets having particularly strong dipolar forces \cite{Compounds, PhysRevB.12.5255, Kotzler1976}.
In particular, the divergence-free nature of the order parameter has been observed using polarized neutron scattering \cite{Kotzler1986}.
See \cite{Gimenez-Grau:2023lpz} for a review.

Surprisingly, there have been few studies of this model using Monte Carlo simulations.
Recently, some of us performed what may have been the first-ever Monte Carlo study \cite{Itou:2025oku}.
In that work, the divergence-free constraint was imposed by introducing a penalty term in the lattice action, causing a significant slowing down in thermalization.
It would be desirable to look for more robust simulation strategies.

\begin{figure}[t]
    \centering
    \includegraphics[width=0.8\linewidth, 
    alt={Graphical explanation of our lattice model and Monte Carlo algorithm.}
    ]{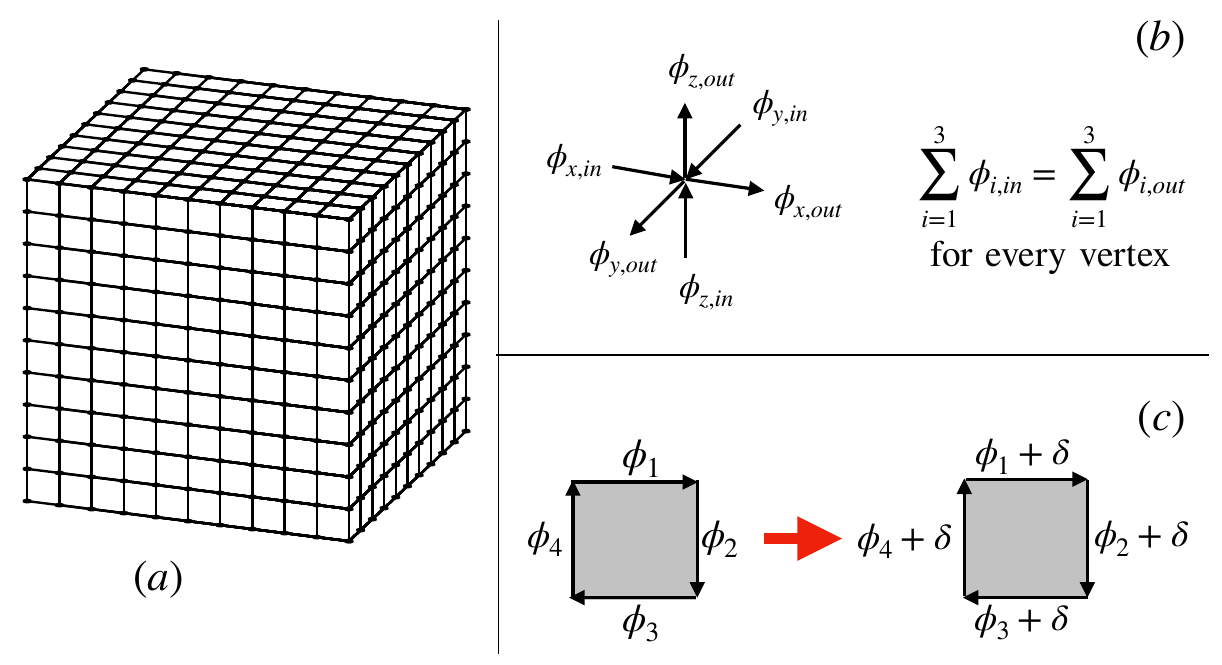}
    \caption{\label{fig:summary} 
        The model and the Monte Carlo algorithm, schematically. 
        See Sections \ref{sec:Model} and \ref{sec:MC} for the precise definition and notation. 
        (a) We will consider a lattice model whose variable will be a vector field flowing along the edges of the cubic lattice. 
        (b) The vector field will satisfy a discrete version of the divergence-free constraint: for every vertex, 
        the sum of inflows along the three directions equals the sum of outflows. 
        (c) The local update increments the values of the vector field 
        flowing along the edges of a randomly chosen plaquette by the same quantity $\delta$. 
        There will also be a global update; see Section \ref{sec:MC}.}
\end{figure}

In this paper, we propose a lattice model expected to lie in the dipolar universality class, which satisfies a discretized version of the divergence-free constraint exactly, rather than due to a penalty term.
We will also formulate a Markov Chain Monte Carlo algorithm that explores the field configuration space ergodically via a combination of Metropolis plaquette updates and a global update for the zero mode.
See Fig.~\ref{fig:summary} for a schematic explanation.
We report results of simulations of our model on the cubic lattice $L\times L \times L$ with periodic boundary conditions up to $L=48$.
The critical exponent $\nu$ of the dipolar universality class is obtained by first-principles calculation for the first time.
We also study whether rotational invariance emerges at the critical point.
Note that the results in Ref.~\cite{Itou:2025oku} were obtained with the finite penalty term, and thus the divergence-free constraint is incomplete.
Taking the limit of the exact constraint is challenging due to the significant slowdown of simulations.
The simulation strategy in this paper avoids this problem by construction.

The paper is structured as follows.
In Section \ref{sec:Model} we will introduce our model, as a lattice discretization of the Aharony-Fisher action and of the divergence-free constraint.
The idea is that the discretized vector field is allowed to flow only along the edges of the cubic lattice, and at every vertex, the total inflow equals the total outflow.
In Section \ref{sec:MC} we describe our Monte Carlo algorithm.
In Section \ref{sec:simulation-results-for-h0} we perform simulations for the most basic version of our model, with a certain quartic coupling $h=0$, which mimics most closely (although not exactly) the rotation invariance of the continuum-limit Aharony-Fisher action.
We see a clear sign of a continuous order-disorder phase transition as a function of the mass parameter.
We locate the phase transition using the Binder ratio and perform the data collapse analysis to determine the critical exponent $\nu$, including the correction-to-scaling term.
We also report the measurements of susceptibility and attempt to determine $\eta$, where only the upper limit is finally given.

We then turn to the study of the rotational invariance at the critical point.
Our lattice model breaks the rotation symmetry to the discrete subgroup of the cubic lattice.
The rotation invariance, however, may emerge at long distances due to a renormalization effect, if the interactions breaking it are irrelevant, or if they turn out to be small accidentally or by tuning.
In Section \ref{sec:q4} we explore this for the $h=0$ model by measuring the anisotropy parameter $Q_4$, whose deviation from the isotropic value turns out to be small.

In Section \ref{sec:q4h} we repeat the analysis for the models with $h=\pm 0.5$.
This confirms the intuition that we can enhance the violation of rotation invariance by playing with this coupling, and also sheds some light on the question of the stability of the rotation-invariant Aharony-Fisher fixed point with respect to the cubic-symmetry-breaking deformation, with some preliminary interpretation in Section \ref{sec:interpretation}.
We conclude in Section \ref{sec:discussion-and-outlook} with a discussion and outlook.

\section{Lattice model \label{sec:Model}}

As shown by Aharony and Fisher \cite{FisherAharony, AharonyFisher, AF-PRBII}, the dipolar phase transition in 3D can be studied using the local Euclidean action\footnote{
We use Einstein's convention with indices $\mu,\nu=1,2,3$.}
\beq
S= \int d^3 x\, [(\partial_\mu \phi_\nu)(\partial_\mu \phi_\nu)+ m^2 (\phi_\mu \phi_\mu)+ g(\phi_\mu \phi_\mu)^2]
\label{eq:3ddip}
\eeq
supplemented by the divergence-free constraint
\beq
\partial_\nu \phi_\nu =0\,.
\label{eq:transverse}
\eeq
Because of the constraint, we can also write the action equivalently as
\beq
S = \int d^3 x\, [\frac 12 \theta_{\mu\nu}\theta_{\mu\nu}+ m^2 (\phi_\mu\phi_\mu)+ g(\phi_\mu\phi_\mu)^2]\,,
\label{eq:3ddip1}
\eeq
where $\theta_{\mu\nu}=\partial_\mu \phi_\nu-\partial_\nu \phi_\mu$ is the "field strength" associated with the "gauge field" $\phi_\mu$.
We will not use this "gauge field" terminology, because gauge invariance is broken by the mass and the quartic terms and by the transverse constraint.
In particular, there is no need to fix the gauge.

In addition to the translation invariance, the symmetry of the continuum model is $O(3)\times \mathbb{Z}_2$ where $O(3)$ corresponds to spatial rotations under which $\phi_\mu$ transforms as a vector, while $\mathbb{Z}_2$ is the internal charge conjugation symmetry $\phi_\mu(x) \to -\phi_\mu(x)$.

\begin{figure}[t]
    \centering
    \includegraphics[width=0.3\linewidth, 
    alt={Figure showing field variables on six edges adjacent to a vertex.}
    ]{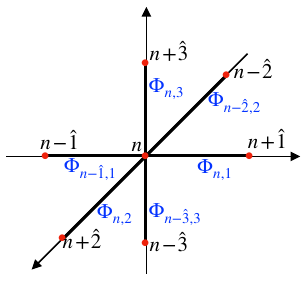}
    \caption{\label{fig:vertex} Lattice variables on 6 edges around a vertex $n$.}
\end{figure}

Our lattice model will be a discretization of \eqref{eq:3ddip1}.
We consider the cubic lattice of size $L\times L\times L$, with periodic boundary conditions.
We label the vertices of the cubic lattice by $n$ and edges coming from the vertex $n$ by $(n,\mu)$ in the positive direction $\mu=1,2,3$.
A lattice vector field is a collection of real fields $\Phi_{n,\mu}$, one per each edge of the lattice.
We stress that on every edge there is just a single lattice field variable $\Phi_{n,\mu}$, with index $\mu$ corresponding to the direction of this edge (see Fig.~\ref{fig:vertex}).
A lattice vector field can be thought of as a discretization of the vector field $\phi_\mu(x)$ appearing in \eqref{eq:3ddip1}, loosely $\Phi_{n,\mu}\sim \phi_\mu(n)$.

We call the lattice vector field divergence-free if it satisfies:
\beq
\sum_{\mu} (\Phi_{n,\mu}-\Phi_{n-\hat{\mu},\mu})=0 \quad \text{for every vertex }n.
\label{eq:lat-transverse}
\eeq
This says that the total inflow into every vertex along all adjacent edges is zero, taking into account the orientation.
This can be thought of as a discretization of the continuum divergence-free constraint \eqref{eq:transverse}.

Our model is defined as follows:
\begin{itemize}
    \item
          The field configuration space consists of lattice vector fields satisfying the divergence-free constraint \eqref{eq:lat-transverse}.
    \item
          The lattice action is a straightforward discretization of \eqref{eq:3ddip1}:
          \beq
          S_{\rm lat} = \sum_{n} \Bigl[\sum_{\mu<\nu} \Theta_{n,\mu\nu}^2+ m^2\sum_{\mu} \Phi_{n,\mu}^2 + \frac g 4 A_{n}^2\Bigr],
          \label{eq:ourmodel}
          \eeq
          where $\Theta_{n,\mu\nu}$ is the "plaquette", while $A_{n}$ is a symmetric discretization of $(\phi_\mu\phi_\mu)(n)$:
          \beq
          \Theta_{n,\mu\nu} = \Phi_{n+\hat{\mu},\nu}  - \Phi_{n,\nu} - \Phi_{n+\hat{\nu},\mu} +\Phi_{n,\mu} \,,
          \eeq
          \beq
          A_n = \sum_{\mu} (\Phi_{n-\hat{\mu},\mu}^2+\Phi_{n,\mu}^2)\,.
          \eeq
\end{itemize}
Our lattice model does not preserve the full rotational $O(3)$ symmetry of \eqref{eq:3ddip1}, but only the discrete subgroup corresponding to the symmetries of the cubic lattice.
Rotational invariance may emerge at the phase transition due to renormalization group (RG) running, if the interactions breaking it are irrelevant and decay away.
It may also emerge accidentally, due to fine-tuning.
This will be studied in Section \ref{sec:q4}.
We will also generalize our model by adding the following extra term to the action:
\beq
h \sum_{n,\mu}  \Phi_{n,\mu}^4\,.
\label{eq:hdef}
\eeq
We will investigate the enhancement of the rotational symmetry breaking caused by the extra coupling $h$ (Section \ref{sec:q4h}).

\begin{remark}
We note that divergence-free vector fields living on links of a lattice appear in other models of statistical physics, notably in frustrated systems exhibiting a Coulomb phase (such as spin ice) \cite{2010ARCMP...1..179H} and in the current representation of the quantum rotor model \cite{PhysRevLett.69.828}.
Our model differs because our link variables take continuous values from $-\infty$ to $+\infty$, whereas in those models they are discrete.
Furthermore, the lattice action is completely different.
\end{remark}

\section{Monte Carlo algorithm \label{sec:MC}}

In order to perform a Monte Carlo sampling that maintains the divergence-free condition \eqref{eq:lat-transverse}, we employ the loop update for the Metropolis algorithm.
In particular, we choose the plaquette update where, given a plaquette $(n,\mu,\nu)$, we update the field variables around the plaquette by
\begin{align}
    \Phi_{n,\mu}         & \;\rightarrow\;\Phi_{n,\mu}+\delta,\nonumber         \\
    \Phi_{n+\hat\mu,\nu} & \;\rightarrow\;\Phi_{n+\hat\mu,\nu}+\delta,\nonumber \\
    \Phi_{n,\nu}         & \;\rightarrow\;\Phi_{n,\nu}-\delta,\nonumber         \\
    \Phi_{n+\hat\nu,\mu} & \;\rightarrow\;\Phi_{n+\hat\nu,\mu}-\delta.
    \label{eq:loop-update}
\end{align}
with $\delta$ uniformly drawn from the range $-\delta_\text{max}<\delta<\delta_\text{max}$.
Once this plaquette update is performed on $L^3$ randomly picked plaquettes, we call it a ``sweep''.

Additionally, for the algorithm to correctly thermalize, we need to ensure that the Markov chain can reach any field configuration satisfying the divergence-free constraint (ergodicity).
This is not possible with just the above plaquette update, which leaves all three components of the magnetization $\sum_n \Phi_{n,\mu}$ invariant.
We need to add three ``global'' updates which can change these. One possibility would be to add three non-contractible loop updates across the three axes \cite{PhysRevLett.69.828}.
Another, equivalent possibility is a uniform global update
\begin{equation}
    \Phi_{n,\mu}\;\rightarrow\;\Phi_{n,\mu} + \delta_\text{global} v_{\mu}.
    \label{eq:global}
\end{equation}
The ergodicity of these updates follows from the mathematical fact that the first simplicial homology group of the three-torus is $H_1(T^3,\mathbb{R})=\mathbb{R}^3$.
Choosing between them is a question of convenience and of optimizing the acceptance rate.
In our work, we used \eqref{eq:global}.
The vector $\vec{v} = (v_{1},v_{2},v_{3})$ was drawn from the unit ball with the probability density of $\rho(\vec{v})=\frac1{4\pi |\vec{v}|^2}$.\footnote{
I.e.~the direction of $\vec{v}$ is uniform on $S^2$ and the length of $\vec{v}$ is uniform in $[0,1]$.}
We stress that $\vec{v}$ is independent of $n$. The global update is performed once after each sweep of plaquette updates.

In principle, $\delta_\text{global}$ and $\delta_\text{max}$ may be optimized to maintain a reasonably high acceptance rate.
However, in this study we chose $\delta_\text{global}=\delta_\text{max}=0.2$ independently of $L$.
For a fixed $\delta_\text{global}$, the acceptance rate of the global update is expected to decrease as $\sim L^{-3/2}$.
This estimate arises from requiring that the quadratic variation of the action in $v$, which at equilibrium scales with the volume as $\sim L^3|\vec{v}|^2$, remains of order one.
Our simulations confirm this $\sim L^{-3/2}$ trend.
Still, even for the largest volumes, the acceptance rate of the global updates was about $7\%$, which was sufficient for our purposes, while for the plaquette updates it was about $75\%$ for all volumes.
In the future, when going to even larger volumes, it makes sense to scale $\delta_\text{global}\sim L^{-3/2}$ to maintain a fixed acceptance rate of the global update.

Additionally, we employed the replica exchange method~\cite{Hukushima:1995rtf, Ilgenfritz:2000nj}.
This is done by running multiple simulations for different $m^2$ in parallel and attempting to swap the configurations after every $N_\text{swap}=50$ sweeps.
We expect that this should reduce autocorrelation time, especially in the ordered phase.

\begin{remark}
    The plaquette update was also used for the simulation of the current representation of the quantum rotor model \cite{PhysRevLett.69.828}. In that model, the divergence-free lattice vector field takes integer values, so they used \eqref{eq:loop-update} with $\delta=\pm1$. As global updates, they used loop updates on loops circling around the periodic lattice in three directions. To alleviate long autocorrelation times, alternative update algorithms, such as the worm algorithm, have been subsequently proposed \cite{2003PhRvE..67a5701A,2003PhRvE..68b6702A}. Because in our problem the field takes continuous values, we simply use global updates of the zero mode with a sufficiently small step to have a sufficiently large acceptance rate. This option was not available to the authors of \cite{PhysRevLett.69.828} since their step size is 1, leading to an exponentially small acceptance rate in their case.
\end{remark}

\section{Simulation results for \texorpdfstring{$h=0$}{h=0} \label{sec:simulation-results-for-h0}}

\subsection{Observables \label{sec:observables}}

To characterize the phase transition and investigate the emergence or breaking of rotational symmetry, we measure several thermodynamic observables during the Monte Carlo simulation. The fundamental quantity of interest is the uniform magnetization vector $\vec{M} = (M_1, M_2, M_3)$, defined as the volume average of the lattice vector field:
\begin{align} M_\mu = \frac{1}{L^3} \sum_{n} \Phi_{n,\mu}, \end{align}
where $\mu \in \{1,2,3\}$. The scalar magnetization $M$ is given by the magnitude of this vector, $M = |\vec{M}|$. Based on the magnetization, we compute the following physical quantities. The magnetic susceptibility $\chi$ is defined as:
\begin{align}
    \chi = L^3 \left( \langle M^2 \rangle - \langle M \rangle^2 \right)\,.
\end{align}

To precisely locate the critical point of the phase transition, we employ the Binder ratio $U$. It is a dimensionless ratio of the moments of the magnetization, defined as
\begin{align}
    U = \frac{5}{2} - \frac{3}{2} \frac{\langle M^4 \rangle}{\langle M^2 \rangle^2
    }\ .
\end{align}
The coefficients in this definition are chosen so that $U$ will tend to $1$ in the ordered phase ($m^2 \ll m^2_c$) while it will tend to $0$ in the disordered phase ($m^2 \gg m^2_c$).\footnote{
The former is obvious.
The latter follows from applying Wick's theorem for three-component Gaussian fluctuations, Gaussianity being expected in the disordered phase.}

Finally, to quantify the breaking of the full $O(3)$ rotational symmetry down to the discrete cubic group of the lattice, we introduce the anisotropy parameter $Q_4$:
\begin{align}
    Q_4 = \frac{\langle \sum_{\mu=1}^3 M_\mu^4 \rangle}{\langle M^4 \rangle} .
    \label{eq:Q4}
\end{align}
This quantity acts as an indicator of rotational invariance. For a system with unbroken $O(3)$ rotational symmetry, the distribution of the magnetization vector $\vec{M}$ is isotropic, which yields $Q_4 = 0.6$.

Deviations from this value indicate that the rotational symmetry is broken, and the system instead exhibits cubic anisotropy.
Deep in the ordered phase, $Q_4$ approaches distinct asymptotic values depending on the symmetry-breaking pattern. If the system undergoes a transition where the symmetry breaks as $O_h \to D_{4h}$ (tetragonal), the magnetization aligns along one of the principal lattice axes, e.g., $\vec{M} \propto (1,0,0)$. In this limit, the anisotropy parameter approaches $Q_4 \to 1$. Conversely, if the symmetry breaks as $O_h \to D_{3d}$ (rhombohedral), the magnetization aligns along the body diagonal of the cubic lattice, e.g., $\vec{M} \propto (1,1,1)$, which yields $Q_4 \to 1/3$. Therefore, the value of $Q_4$ distinguishes the nature of the symmetry breaking: $Q_4 > 0.6$ indicates an axis-aligned (tetragonal) ordering, while $Q_4 < 0.6$ signifies a diagonal-aligned (rhombohedral) ordering.

\subsection{Scaling analysis and critical point \label{sec:Binder}}

The critical point $m^2_c$ and critical exponent $\nu$ of the phase transition can be estimated by the data collapse analysis, which utilizes the finite-size scaling of observables.
In this section, we briefly describe our data-collapse method for the Binder ratio $U$ and present the results.
We used the Monte Carlo data set of $U$ obtained for various $m^2$ and $L$ selected from the range $-0.719 \leq m^2 \leq -0.700$ near the critical point $m_c^2 \sim -0.71$ and $10 \leq L \leq 48$, while fixing $g = 1.0$.
For each parameter, the simulation was performed over $10^8$ sweeps (following $10^4$ thermalization sweeps), with measurements taken every $100$ sweeps.
The expectation value $U(m^2, L)$ and the statistical error $\Delta_U(m^2, L)$ were estimated by the jackknife method.

\begin{figure}[t]
    \centering
    \includegraphics[width=0.6\linewidth, 
    alt={Graph depicting the Binder ratio near the critical point for various system sizes.}
    ]{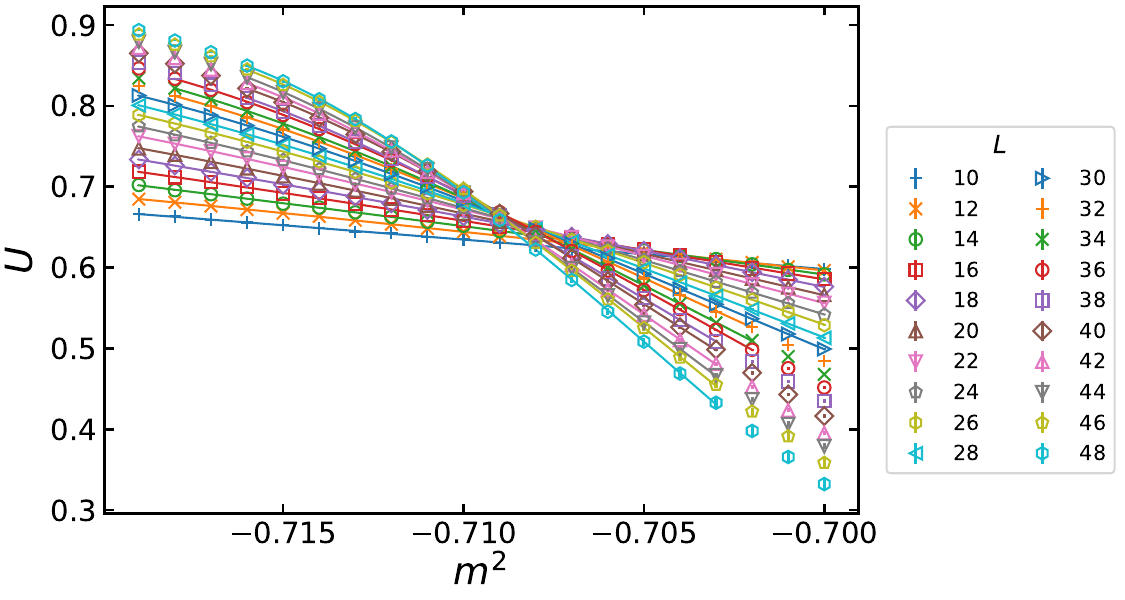}
    \caption{\label{fig_Ufit_vs_m2} 
        The Binder ratio $U$ is plotted against $m^2$ for various $L$. 
        The solid curves are the results of fitting with the interpolation function \eqref{eq_interpolation_U} 
        and are drawn only inside the fitting range.}
\end{figure}

First, we obtain the interpolation function $\tilde{U}_L(m^2)$ of the Binder ratio for each $L$ for later use.
Here we adopt the cubic polynomial of $m^2$,
\begin{equation}
    \tilde{U}_{L}(m^{2})=c_{0}(L)+c_{1}(L)\,m^{2}+c_{2}(L)\,m^{4}+c_{3}(L)\,m^{6}.
    \label{eq_interpolation_U}
\end{equation}
The coefficients $\{c_{i}(L)|i=0,1,2,3\}$ are determined by fitting the data points of $U(m^2, L)$ by the polynomial for each $L$.
The fitting range $\mathcal{R}(L)=[m_{\mathrm{min}}^2, \, m_{\mathrm{max}}^2]$ is chosen so that the deviations of the data points from the fitting curve are comparable to the statistical error and not biased.
The Monte Carlo results of $U$ and the fitting curves are shown in Fig.~\ref{fig_Ufit_vs_m2}.

The data collapse analysis is based on the scaling behavior of the observable near the critical point $m_c^2$, where the $m^2$ dependence of the observable across different system sizes is characterized by a universal function after an appropriate rescaling.
Here we consider the scaling ansatz of $U$ including a finite size correction $L^{-\omega}$:
\begin{equation}
    U = F\left[ L^{1/\nu}\frac{m^2 - m_c^2}{|m_c^2|} \right] + c_U L^{-\omega},
    \label{eq_dc_ansatz_A}
\end{equation}
where $F[\cdots]$ represents an arbitrary smooth function \cite{Binder:1981sa}.
We will refer to this type of ansatz as {\it ansatz A}.
From the RG perspective, the "correction-to-scaling" term $c_U L^{-\omega}$ accounts for the rate of approach to the fixed point with the increase of the lattice size.
The correction-to-scaling exponent $\omega$ is related by $\omega=\Delta-d$ with $d=3$ to the scaling dimension $\Delta$ of the leading irrelevant operator.
We have four free parameters, $m_c^2$, $\nu$, $\omega$, and $c_U$, which are to be optimized.
When the parameters are set to the optimal values, the data points of $U(m^2, L)$ will collapse into a single curve $Y=F[X]$ on the $(X, Y)$ plane by rescaling
\begin{equation}
    X = L^{1/\nu}\frac{m^2 - m_c^2}{|m_c^2|}, \quad Y = U(m^{2},L) - c_U L^{-\omega}.
    \label{eq_rescaled_axes_A}
\end{equation}

Next, we introduce a cost function for optimizing the parameters, which measures the quality of the data collapse.
Based on Ref. \cite{Bhattacharjee_2001}, the cost function $P$ for ansatz A is defined as
\begin{align}
    P:= & \frac{1}{2N_{\mathrm{over}}} \sum_L \sum_{L_0(\neq L)} \sum_{m^2\in\mathcal{M}(L;L_0)} W(m^2,L)^2 \nonumber \\
        & \times \left| U(m^2,L) - c_U L^{-\omega} \vphantom{\tilde{U}_{L_0}\left(m_{\mathrm{over}}^2 (L_0;m^2,L)\right)} 
        - \tilde{U}_{L_0}\left(m_{\mathrm{over}}^2 (L_0;m^2,L)\right) + c_U L_0^{-\omega} \right|^2.
    \label{eq_cost_function}
\end{align}
The norm $|\cdots|^2$ is the squared distance from the rescaled data point $Y$ for $(m^2, L)$ to the rescaled interpolation function $\tilde{U}_{L_0}$ for $L_0 \neq L$.
Here $m_{\mathrm{over}}^2(L_0; m^2, L)$ represents a solution of
\begin{equation}
    L^{1/\nu}\frac{m^2 - m_c^2}{|m_c^2|} = L_0^{1/\nu}\frac{m_{\mathrm{over}}^2 - m_c^2}{|m_c^2|},
    \label{eq_overlap_m2}
\end{equation}
which indicates that $m^2$ and $m_{\mathrm{over}}^2$ fall into the same horizontal position $X$ after rescaling.
Note that the unknown smooth function $F[\cdots]$ is now approximated by the polynomial interpolation function \eqref{eq_interpolation_U} within the fitting range $\mathcal{R}(L)$.

The squared distance is weighted by the statistical error $\Delta_{U}(m^2, L)$ of the data point, and thus the weight factor is
\begin{equation}
    W(m^{2},L) = \frac{1}{\Delta_{U}(m^{2},L)}.
\end{equation}
The sum over $m^2$ in Eq.~\eqref{eq_cost_function} is restricted to the so-called overlap region $\mathcal{M}(L; L_0)$ defined by 
\begin{align}
    \mathcal{M}(L; L_0) = 
    & \left\{ m^2 \in\mathcal{R}(L) \middle| \exists m_0^2 \in\mathcal{R}(L_0), 
    L^{1/\nu}\frac{m^2 - m_c^2}{|m_c^2|} = L_0^{1/\nu}\frac{m_0^2 - m_c^2}{|m_c^2|} \right\}.
\end{align}
If the solution $m_{\mathrm{over}}^2(L_0; m^2, L)$ of Eq.~\eqref{eq_overlap_m2} does not exist inside the fitting range $\mathcal{R}(L_0)$, the data point is excluded from the sum.
Finally, the cost function is normalized by the number of data points included in the sum,\footnote{
    When $N_{\mathrm{over}} \gg 1$ and the statistical error of the interpolated value $ \tilde{U}_{L_0}(m_{\mathrm{over}}^2)$ is approximately equal to $\Delta_{U}(m^2, L)$, our definition of $P$ is equivalent to the chi-square cost functions in Refs.~\cite{doi:10.1143/JPSJ.62.435, Houdayer_2004}.}
\begin{equation}
    N_{\mathrm{over}}=\sum_{L_{0}}\sum_{L(\neq L_{0})}\sum_{m^{2}\in\mathcal{M}(L_{0};L)}1.
\end{equation}
The data collapse was performed by minimizing the cost function $P$ using the numerical optimization method (trust-region reflective algorithm).
The location of the minimum in the 4-dimensional parameter space is then the estimate of $m_c^2$, $\nu$, $\omega$, and $c_U$.
The jackknife method is applied to the entire data collapse analysis to obtain the statistical error.

\begin{figure}[t]
    \centering
    \includegraphics[width=0.4\linewidth, 
    alt={Graph depicting the Binder ratio for various system sizes rescaled with ansatz A, collapsing into a single curve.}
    ]{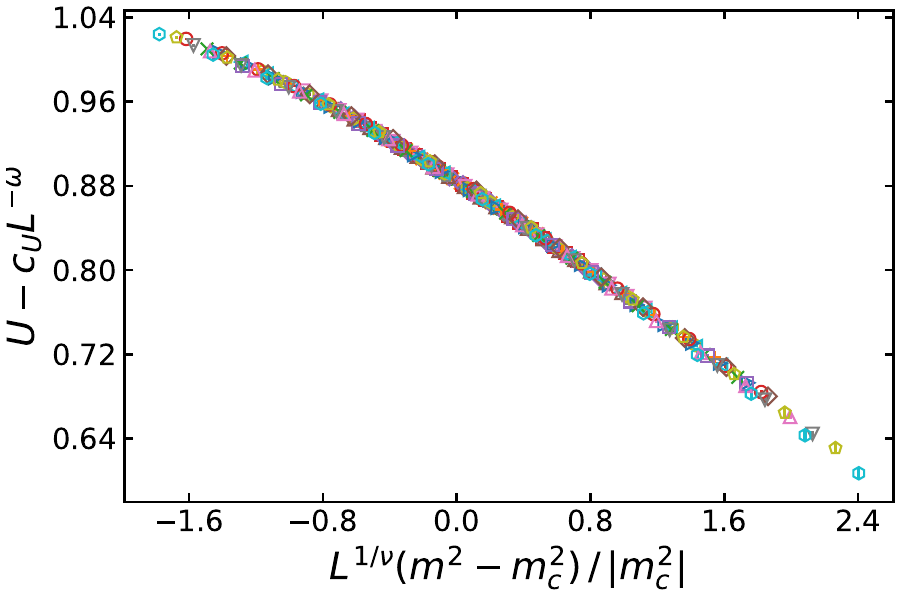}
    \includegraphics[width=0.5\linewidth, 
    alt={Graph depicting the Binder ratio for various system sizes rescaled with ansatz B, collapsing into a single curve.}
    ]{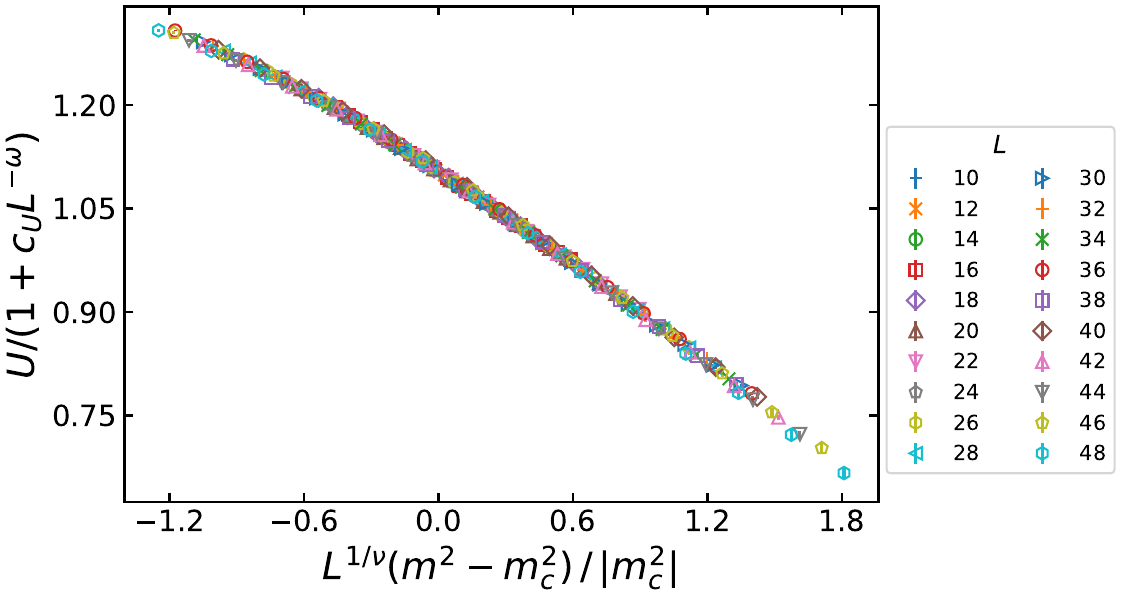}
    \caption{\label{fig_data_collapse_U} 
        The left and right panels show the results of data collapse analysis with ansatzes A and B, respectively. 
        The Binder ratio is plotted on the rescaled axes $(X, Y)$ with the corresponding ansatz, 
        \eqref{eq_rescaled_axes_A} or \eqref{eq_rescaled_axes_B}, for various $L$.}
\end{figure}

Let us show the results of the data collapse analysis with ansatz A \eqref{eq_dc_ansatz_A}.
The optimized parameters are summarized in the middle row of Table \ref{tab_dc_U} with the corresponding value of the cost function $P$.
The table also shows the value of the function $F$ in ansatz A at the critical point $m^2 = m^2_c$, which is obtained by averaging $F[0] = \tilde{U}_L(m^2_c) - c_U L^{-\omega}$ over all $L$.
We rescale $m^2$ and $U(m^2, L)$ to $X$ and $Y$ as in Eq.~\eqref{eq_rescaled_axes_A} using these parameters and plot the resulting data points $(X, Y)$ in the left panel of Fig.~\ref{fig_data_collapse_U}.
The plot shows that the data points collapse into a single curve as expected.

\begin{table}[t]
    \centering
    \begin{tabular}{|c||c|c|c|c|c|c|}
        \hline
        ansatz & $m^2_c$      & $\nu$      & $\omega$  & $c_U$      & $F[0]$    & $P$ \tabularnewline
        \hline \hline
        A      & $-0.7105(1)$ & $0.713(3)$ & $0.23(3)$ & $-0.42(3)$ & $0.88(4)$ & $2.04$ \tabularnewline
        \hline
        B      & $-0.7107(1)$ & $0.756(5)$ & $0.12(3)$ & $-0.56(4)$ & $1.1(1)$  & $1.95$ \tabularnewline
        \hline
    \end{tabular}
    \caption{\label{tab_dc_U}
        The optimized parameters obtained by the data collapse with ansatzes A and B are listed
        with the corresponding $F[0]$ value and cost function $P$.}
\end{table}

So far, we have used ansatz A \eqref{eq_dc_ansatz_A} for the data collapse.
However, the form of the ansatz cannot be uniquely determined from a theoretical perspective, and there are multiple valid alternatives.
Thus, uncertainty in the ansatz leads to systematic errors in the estimated critical point and exponent.
To evaluate the systematic error, we consider another type of ansatz,
\begin{equation}
    U = (1 + c_U L^{-\omega}) F\left[ L^{1/\nu}\frac{m^2 - m_c^2}{|m_c^2|} \right],
    \label{eq_dc_ansatz_B}
\end{equation}
which is called {\it ansatz B}.
The number of free parameters is the same as in ansatz A.
We adjusted the cost function $P$ for ansatz B and performed the data collapse analysis again.
While the rescaling of the $m^2$ axis is the same as $X$ of Eq.~\eqref{eq_rescaled_axes_A}, the rescaling of $U$ is modified as
\begin{equation}
    Y = \frac{U(m^{2},L)}{1 + c_U L^{-\omega}}.
    \label{eq_rescaled_axes_B}
\end{equation}
The optimized parameters are summarized in the bottom row of Table \ref{tab_dc_U}, and the rescaled data points are shown in the right panel of Fig.~\ref{fig_data_collapse_U}.
The collapse of the data points is observed for this case as well.
Since both ansatzes A and B give almost the same value of the cost function, $P \sim 2$, they are considered equally reasonable ansatzes.
Indeed, the results of $m^2_c$ are consistent with each other.
On the other hand, the results of $\nu$ exhibit a non-negligible discrepancy, which indicates the existence of large systematic uncertainty.

From the RG perspective, a more theoretically justified fitting ansatz would be $U = F[x]+L^{-\omega} G[x]$ with $x=L^{1/\nu}\frac{m^2 - m_c^2}{|m_c^2|}$.
However, the current amount of data is insufficient to fit two independent functions $F$ and $G$.
We thus used ansatzes A and B as proxies, which have $G$ as a constant and $G \propto F$, respectively.
Although we can include further subleading correction terms in the ansatz, adding more parameters typically results in overfitting; namely, the cost function has flat directions and is unable to determine the parameters.
We observed that such further subleading corrections yield no more improvement in the quality of data collapse.
Thus, we consider only ansatzes A and B to compare the data-collapse results and evaluate the systematic error.

\begin{figure}[t]
    \centering
    \includegraphics[width=0.9\linewidth, 
    alt={Graph depicting the dependence of data-collapse results on the system-size cutoff, which is negligible within the statistical errors.}
    ]{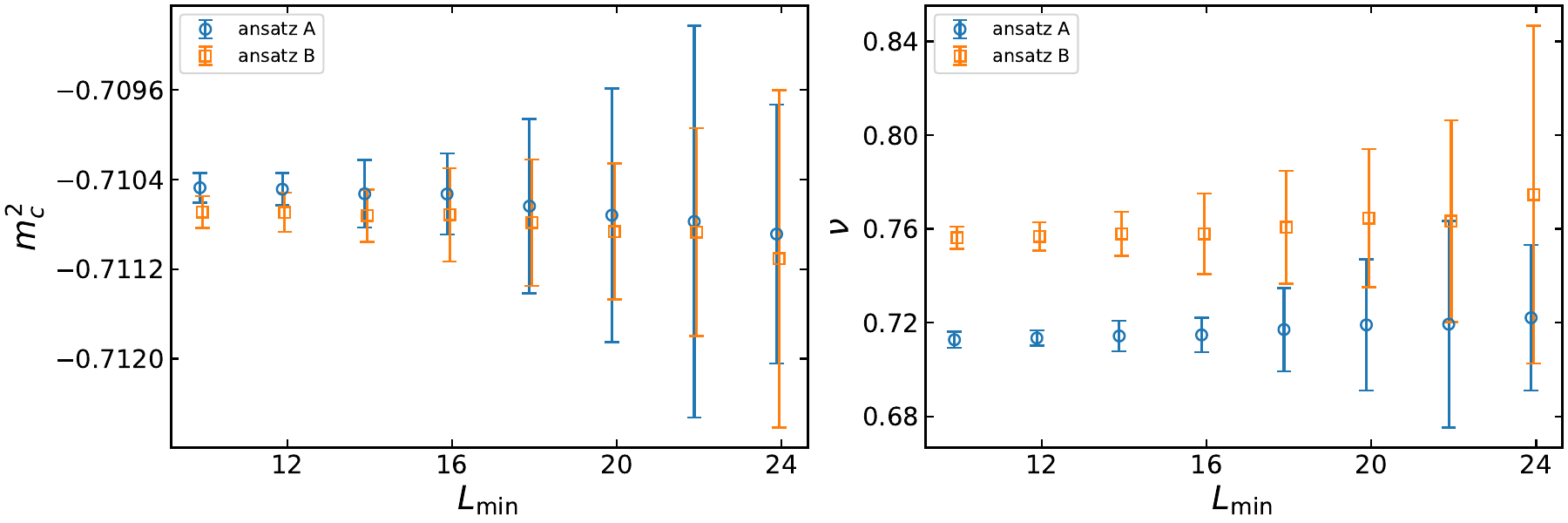}
    \caption{\label{fig_critical_vs_Lmin} 
        The data collapse results of the critical point $m^2_c$ (left) 
        and critical exponent $\nu$ (right) are plotted against the cutoff $L_\mathrm{min}$ for ansatzes A and B.}
\end{figure}

To examine the validity of the finite-size corrections, we also inspected the dependence on the fitting range of $L$.
We removed the data points of small $L$ by setting the cutoff as $L_\mathrm{min} \leq L$ and performed the data collapse for various $L_\mathrm{min}$.
Figure \ref{fig_critical_vs_Lmin} shows the $L_\mathrm{min}$ dependence of the resulting $m^2_c$ and $\nu$ for ansatzes A and B.
Note that the sizes of the error bars increase with the cutoff $L_\mathrm{min}$ since the amount of data decreases with $L_\mathrm{min}$.
It is confirmed that the results without the cutoff ($L_\mathrm{min} = 10$) agree with those with the cutoff ($L_\mathrm{min} > 10$).
Both ansatzes provide the finite-size correction consistently in a wide range of $L$ and do not cause additional systematic error from uncertainty in the fitting range.
Therefore, we obtained the final estimates by simply averaging the results of the two ansatzes at $L_\mathrm{min} = 10$ as 
\begin{align}
    & m^2_c = -0.71058 (14)_\mathrm{stat} (11)_\mathrm{sys}, \\
    & \nu = 0.735 (5)_\mathrm{stat} (22)_\mathrm{sys},
\end{align}
where the statistical error is given by the larger of those for the two results, and the systematic error is half of the discrepancy.
Although there exists a large systematic uncertainty from the choice of ansatz, this is the first estimation of the exponent $\nu$ of the dipolar universality class by the Monte Carlo simulation with the exact divergence-free constraint.

\subsection{Susceptibility \label{sec:susceptibility}}

We now turn our attention to the finite-size scaling analysis of the magnetic susceptibility $\chi$, with the motivation to extract $\eta$.
It is widely recognized that estimating critical exponents from susceptibility data is inherently more challenging and less reliable than using the Binder ratio $U$.
To mitigate this issue, we will perform a simultaneous data collapse of both $U$ and $\chi$.
This joint fitting procedure is also expected to refine our estimates for the correlation length exponent $\nu$.

Following the procedure described in Section \ref{sec:Binder}, we employ two distinct scaling ansatzes: ansatz A is defined as
\begin{align}
    U    & = F\left[ L^{1/\nu} \frac{m^2 - m_c^2}{|m_c|^2}\right] + c_U L^{-\omega},                                \\
    \chi & = L^{2-\eta} \left( H\left[ L^{1/\nu} \frac{m^2 - m_c^2}{|m_c|^2} \right] + c_\chi L^{-\omega} \right) ,
\end{align}
while ansatz B is given by
\begin{align}
    U    & =  (1 + c_U L^{-\omega}) F\left[ L^{1/\nu} \frac{m^2 - m_c^2}{|m_c|^2} \right],             \\
    \chi & = (1+c_\chi L^{-\omega}) L^{2-\eta} H\left[ L^{1/\nu} \frac{m^2 - m_c^2}{|m_c|^2} \right] .
\end{align}
In both cases, we assume that the correction-to-scaling exponent $\omega$ is common.

\begin{figure}[t]
    \centering
    \includegraphics[width=0.6\linewidth, 
    alt={Graph depicting the magnetic susceptibility near the critical point for various system sizes.}
    ]{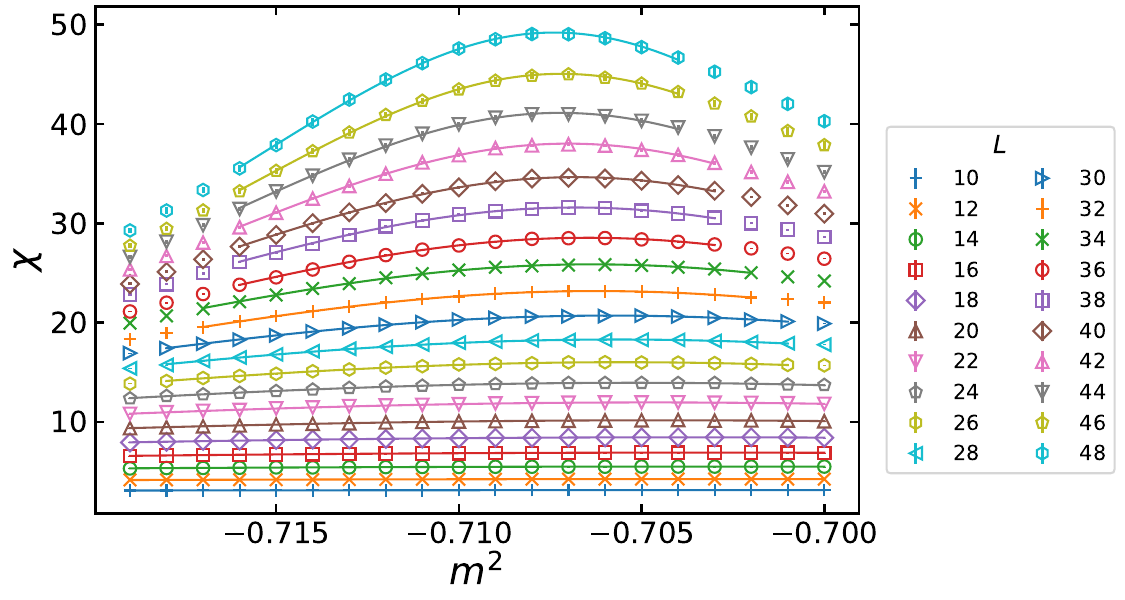}
    \caption{\label{fig_chifit_vs_m2} 
        The magnetic susceptibility $\chi$ is plotted against $m^2$ for various $L$. 
        The solid curves are the results of fitting with the interpolation function \eqref{eq_interpolation_chi} 
        and are drawn only inside the fitting range.}
\end{figure}

Similarly to the case of the Binder ratio $U$, the susceptibility $\chi$ for each $L$ is interpolated using the Lorentzian function of $m^2$,
\begin{equation}
    \tilde{\chi}_{L}(m^2) = \frac{d_0}{ ( m^2 - d_1 )^2 + d_2},
    \label{eq_interpolation_chi}
\end{equation}
where the coefficients $\{d_i(L)|i=0,1,2\}$ are determined by fitting the data points.
The Monte Carlo results of $\chi$ and the fitting curves are shown in Fig.~\ref{fig_chifit_vs_m2}.

\begin{figure}[t]
    \centering
    \includegraphics[width=0.425\linewidth, 
    alt={Graph depicting the joint-data-collapse results with ansatz A.}
    ]{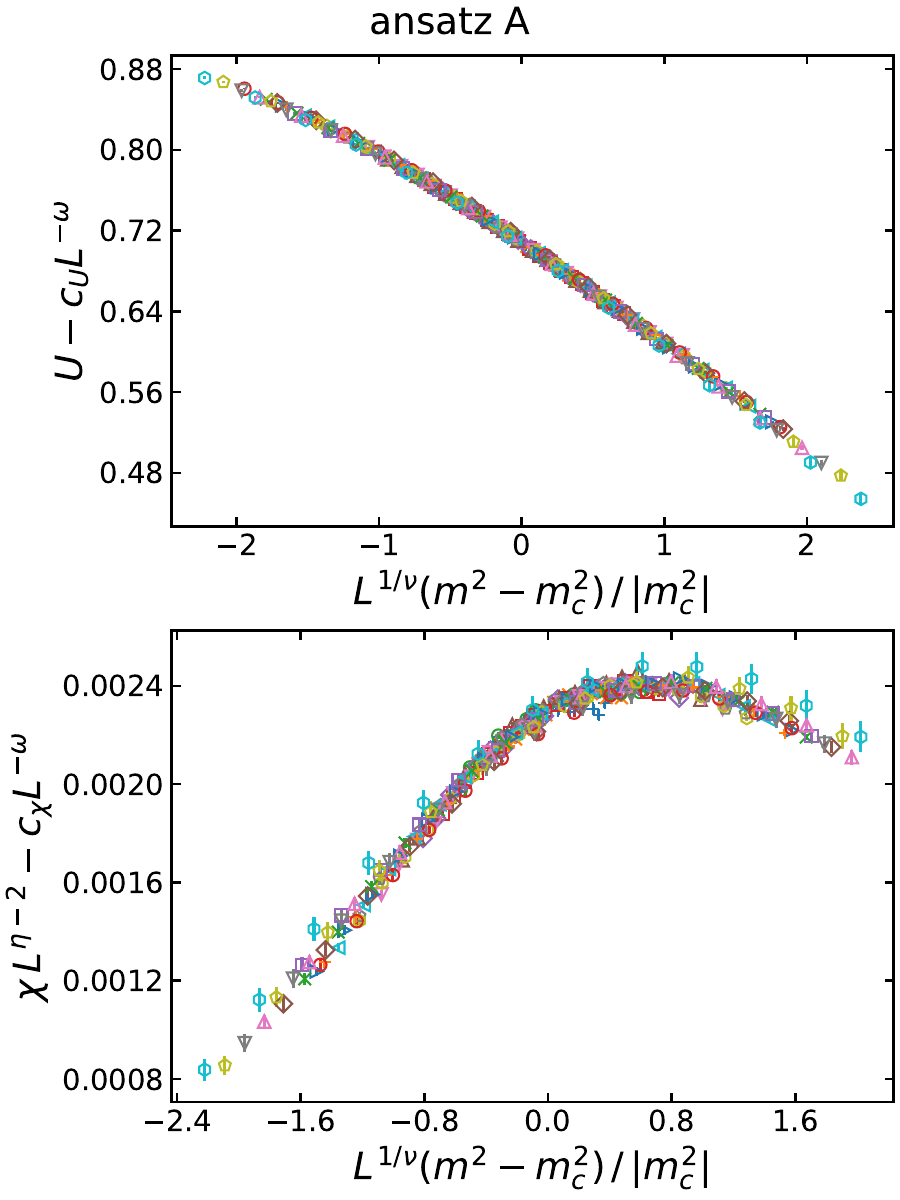}
    \includegraphics[width=0.475\linewidth, 
    alt={Graph depicting the joint-data-collapse results with ansatz B.}
    ]{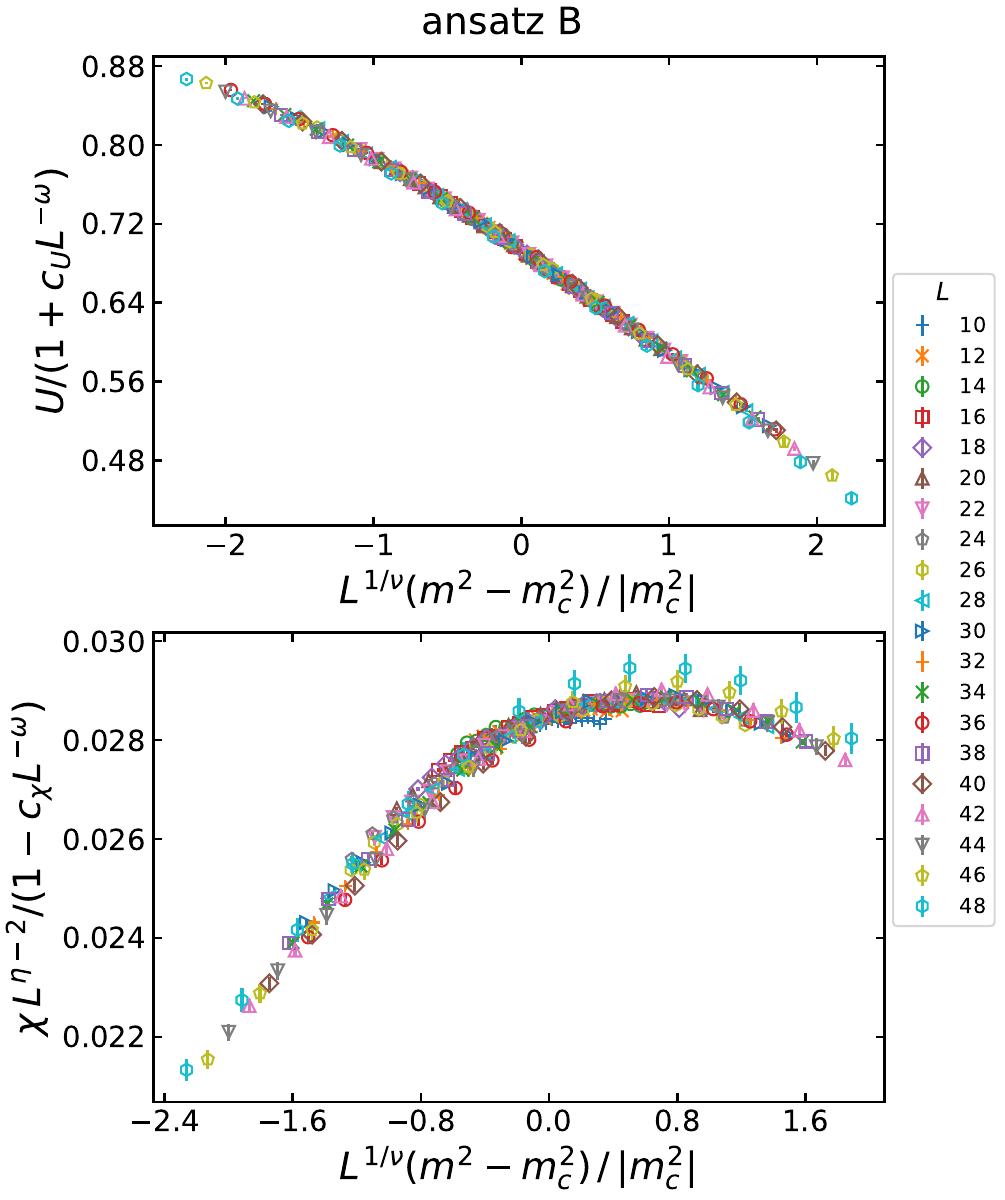}
    \caption{\label{fig_joint_collapse} 
        The results of the joint data-collapse analyses with ansatzes A and B are shown in the left and right columns, respectively, 
        where the Binder ratio $U$ (top) and magnetic susceptibility $\chi$ (bottom) are plotted on the rescaled axes.}
\end{figure}

The global cost function $P$ to be minimized is constructed as the average of the individual cost functions for $U$ and $\chi$.\footnote{
    Since the number of data points used in each fit is close (i.e., $N_U = 355$ and $N_\chi = 341$), 
    this $P$ is more or less equivalent to the reduced $\chi^2$ of the joint fitting.}
The optimal fitting parameters obtained from this procedure are summarized in Table \ref{tab_dc_chi}, and the resulting data collapses for ansatzes A and B are illustrated in the left and right columns of Fig.~\ref{fig_joint_collapse}, respectively.
Visually, the quality of the collapse for $\chi$ is inferior to that of $U$.
This corroborates the empirical observation that the susceptibility is sensitive to statistical and systematic uncertainties, making it more difficult to extract critical exponents than from the Binder ratio.

\begin{table}[t]
    \centering
    \begin{tabular}{|c|c|c|c|c|c|c|c|c|}
        \hline
        ansatz & $m^2_c$       & $\nu$     & $\eta$     & $\omega$  & $c_U$      & $c_\chi$   & $F[0]$     & $P$ \tabularnewline
        \hline \hline
        A      & $-0.70972(9)$ & $0.70(1)$ & $-0.33(7)$ & $0.81(5)$ & $-0.49(4)$ & $0.079(3)$ & $0.708(4)$ & $3.10$ \tabularnewline
        \hline
        B      & $-0.7095(1)$  & $0.70(1)$ & $0.11(1)$  & $0.95(6)$ & $-0.78(7)$ & $3.6(1)$   & $0.693(5)$ & $9.11$ \tabularnewline
        \hline
    \end{tabular}
    \caption{\label{tab_dc_chi}
        The optimized parameters obtained by the data collapse using ansatzes A and B, along with their corresponding $F[0]$ value and cost function $P$.}
\end{table}

\begin{figure}[t]
    \centering
    \includegraphics[width=0.45\linewidth, 
    alt={Graph depicting the cost-function landscape for the joint data collapse with ansatz A, showing a broad valley structure.}
    ]{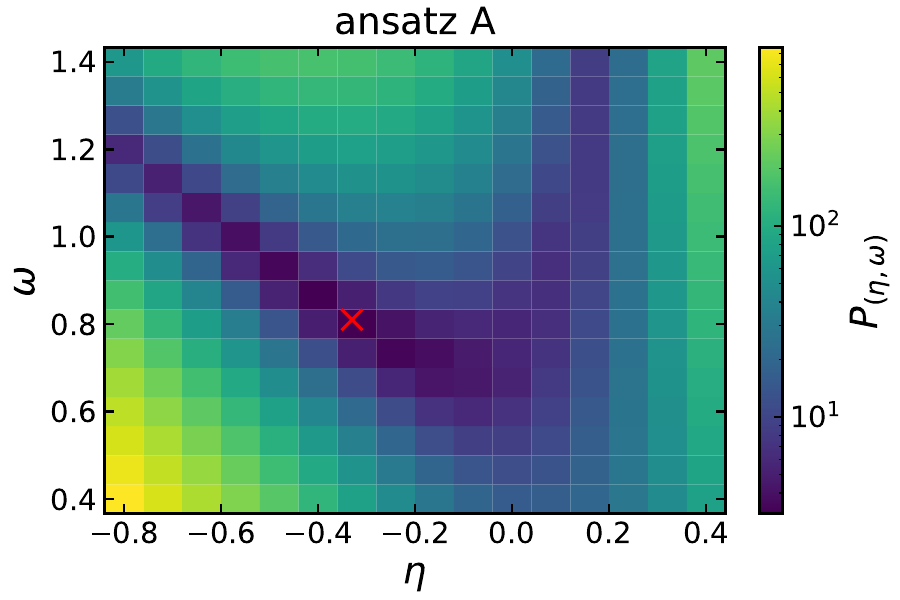}
    \includegraphics[width=0.45\linewidth, 
    alt={Graph depicting the cost-function landscape for the joint data collapse with ansatz B, showing a broad valley structure.}
    ]{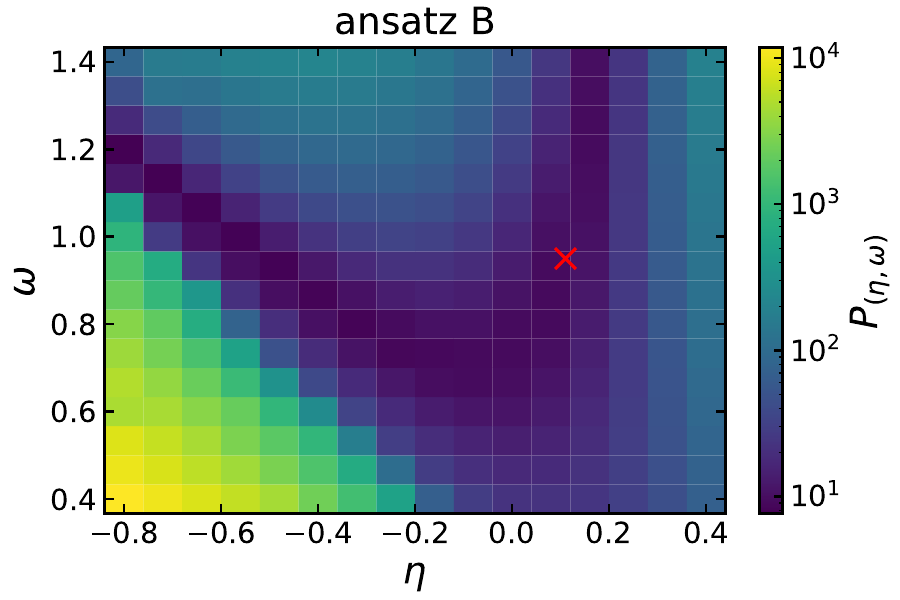}
    \caption{\label{fig_landscape} 
        The cost function $P$ for the joint data collapse with fixed $\eta$ and $\omega$ is plotted as a log-scaled color map on the $(\eta, \omega)$ plane.
        The left and right panels correspond to ansatzes A and B, respectively.
        The parameters other than $\eta$ and $\omega$ are selected to minimize $P$.
        The red markers indicate the optimized parameters in Table \ref{tab_dc_chi}.}
\end{figure}

We observed that the determination of $\eta$ is unstable, and the $\eta$ values obtained from the two ansatzes show a large discrepancy.
In particular, ansatz A yields a negative value while the other is positive.
This instability originates partially from the landscape of the cost function $P$ being flat along the $\omega$ direction, and the determination of $\eta$ is sensitive to the value of $\omega$.
The landscapes of $P$ are shown in Fig.~\ref{fig_landscape}, which have broad, flat regions.
More precisely, a theoretically justified ansatz would be $\chi = L^{2-\eta}(H[x] + L^{-\omega} I[x])$.
However, this ansatz has an intrinsic degeneracy in the fitting: 
$L^{2-\eta}(H[x] + L^{-\omega} I[x]) = L^{2-\eta-\omega}(I[x] + L^{+\omega} H(x))$ if negative $\eta$ and $\omega$ are allowed.
While ansatz A artificially lifts this degeneracy, ansatz B retains it.
Note that the optimized value of $\omega$ also deviates from the one obtained in Section \ref{sec:Binder} using only the Binder ratio.
Physically, this discrepancy suggests that the susceptibility is dominated by different, potentially higher-order, sources of corrections to scaling.

Another source of instability is the presence of spurious local minima in the cost function landscape.
In these potentially unphysical minima, the leading constant term in the scaling function $H(0)$ vanishes, causing the leading correction term to erroneously dictate the main scaling behavior.
This scenario typically yields an unphysical negative value for $\eta$, and the best-fit value with ansatz A is affected by this issue.
As discussed later in Section \ref{sec:interpretation}, the dipolar fixed point may be unstable against cubic perturbations, implying the presence of an additional relevant direction unprotected by lattice symmetries.
In such a scenario, a negative $\eta$ accompanied by a non-zero but small constant term in $H$ could actually have a physical interpretation; 
it is equivalent to the scaling ansatz with positive $\eta$ but with negative $\omega$.
Eventually, we here limit ourselves to putting a weak bound on the estimate of $\eta$ as our Monte Carlo result, 
\begin{equation}
    \eta \lesssim 0.2.
\end{equation}
This limit is compatible with the flat region of the landscape in Fig.~\ref{fig_landscape}.
As for $m^2_c$ and $\nu$, the results from ansatzes A and B in Table \ref{tab_dc_chi} are consistent.
We extract the estimates for them by averaging the two results.
Since there is no a priori theoretical reason to prefer one ansatz over the other, we treat the two ansatzes with equal weight to avoid artificial bias.
The systematic errors are conservatively estimated based on the half-difference between them, while the statistical errors are given by the uncertainties of the individual fits;
\begin{align}
    & m^2_c = -0.70959 (11)_\mathrm{stat} (13)_\mathrm{sys}, \\
    & \nu = 0.702 (12)_\mathrm{stat} (1)_\mathrm{sys}.
\end{align}
Encouragingly, our estimate for $m^2_c$ and $\nu$ remains robust against these instabilities, further validating the reliability of the determination made in the preceding section.

We highlight a notable improvement of this joint fitting procedure:
The estimated value of the critical Binder parameter, $U(m_c^2, L= \infty) = F[0]$, becomes more stable and physically sound compared to the estimates obtained in Section~\ref{sec:Binder}.
In the previous Binder fit, ansatz B yielded an unphysical value of $F \approx 1.1$, which violates the mathematical upper bound of $U \le 1$ dictated by the Cauchy-Schwarz inequality, $\langle M^4 \rangle \ge \langle M^2 \rangle^2$.
In contrast, the joint fit drives the optimized value of $\omega$ to be much larger, suppressing the finite-size correction to $F[0]$ and bringing it down to a stable value of $F \approx 0.7$ for both ansatze.

\subsection{Cubic anisotropy \label{sec:q4}}

Unlike the Heisenberg model, the internal and spatial degrees of freedom in our model are tightly coupled. In the Aharony-Fisher theory \eqref{eq:3ddip}, the divergence-free constraint $\partial_\mu \phi_\mu = 0$ breaks the independent spatial and internal $O(3)$ symmetries down to their diagonal subgroup, corresponding to simultaneous rotations of both the spatial coordinates and the vector field components.

In our lattice action $S_{\rm lat}$ with the discrete transverse constraint \eqref{eq:lat-transverse}, this diagonal $O(3)$ symmetry is explicitly broken down to the discrete cubic point group $O_h$. Our lattice model is invariant under the following symmetry:

\begin{enumerate}

    \item Lattice translations: Invariance under shifts by lattice vectors, subject to periodic boundary conditions.

    \item Cubic rotations and reflections: Invariance under the $90^\circ$ rotations and mirror reflections of the cubic lattice, provided that the spatial indices $\mu$ of the link variables $\Phi_{n,\mu}$ are simultaneously permuted and flipped to match the rotated link orientations.

    \item Global $\mathbb{Z}_2$ symmetry: Invariance under the global sign flip $\Phi_{n,\mu} \to -\Phi_{n,\mu}$ for all $n$ and $\mu$. Note that this is an internal symmetry independent of spatial parity since the spatial coordinates $n$ and the link orientations $\mu$ remain unchanged. It corresponds to the charge conjugation $\mathbb{Z}_2$ symmetry of the continuum model (see Section \ref{sec:Model}).
\end{enumerate}

Because the lattice discretization reduces the continuous rotational symmetry to the discrete $O_h$ group, it is crucial to investigate whether the observed second-order phase transition recovers the full rotational invariance in the continuum limit or shows the cubic instability.\footnote{
The restoration of this continuous rotational symmetry relies on whether the higher-rank tensor operators that break the $O(3)$ symmetry down to the discrete cubic group are irrelevant.
Because the dipolar universality class lacks conformal invariance and reflection positivity, we cannot rely on unitarity bounds of CFTs to guarantee this restoration.}

\begin{figure}[t]
    \centering
    \includegraphics[width=0.9\linewidth, 
    alt={Graph depicting the angular distribution of the magnetization vector, showing nonuniform distribution in the disordered phase.}
    ]{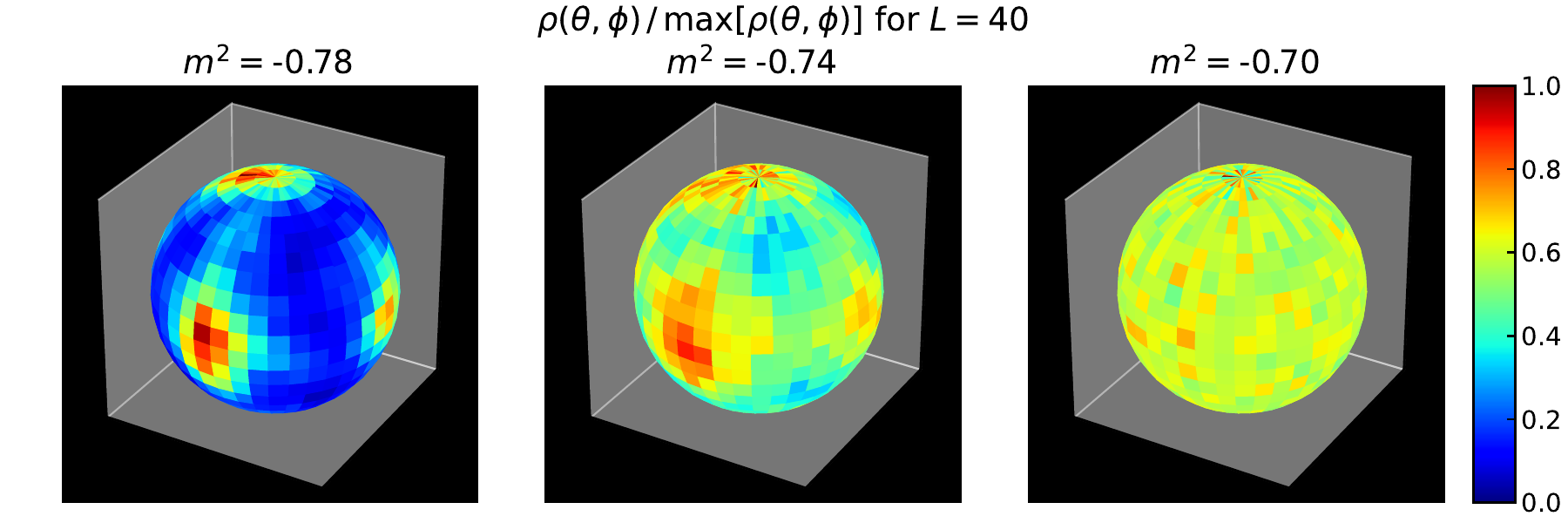}
    \caption{\label{fig_distribution_M_3d} 
        The angular distribution $\rho(\theta,\phi)$ of the magnetization vector $\vec{M}$ is plotted as a colormap on a sphere 
        for $m^2 = -0.78$ (left), $-0.74$ (center), and $-0.70$ (right) with the lattice size $L = 40$. 
        The distribution is normalized to the 0-1 range for use with the colormap.}
\end{figure}

To this end, we first examine the angular distribution $\rho(\theta,\phi)$ of the magnetization vector $\vec{M}$,
where $\theta = \arccos( M_3 / |\vec{M}| )$ and $\phi = \arctan( M_2 / M_1 )$.\footnote{
This definition is meaningful even in the disordered phase, since every sample has a nonzero $\vec{M}$.}
This is shown in Fig.~\ref{fig_distribution_M_3d} for three different values of $m^2$: two below the critical point and one slightly above (recall $m_c^2\approx -0.71$).
In the left panel, deep in the ordered phase ($m^2 \ll m_c^2$), the distribution clearly breaks the rotational symmetry, favoring a tetragonal alignment along the principal axes (see the discussion at the end of Section \ref{sec:MC}). However, as the system gets closer to the critical point (center panel), or goes slightly into the disordered phase (right panel), the distribution becomes increasingly isotropic, suggesting a restoration of the $O(3)$ symmetry at the critical point.

\begin{figure}[t]
    \centering
    \includegraphics[width=0.45\linewidth, 
    alt={Graph depicting the anisotropy parameter near the critical point, showing that the sub-lattice results are closer to the isotropic limit than the entire-lattice ones.}
    ]{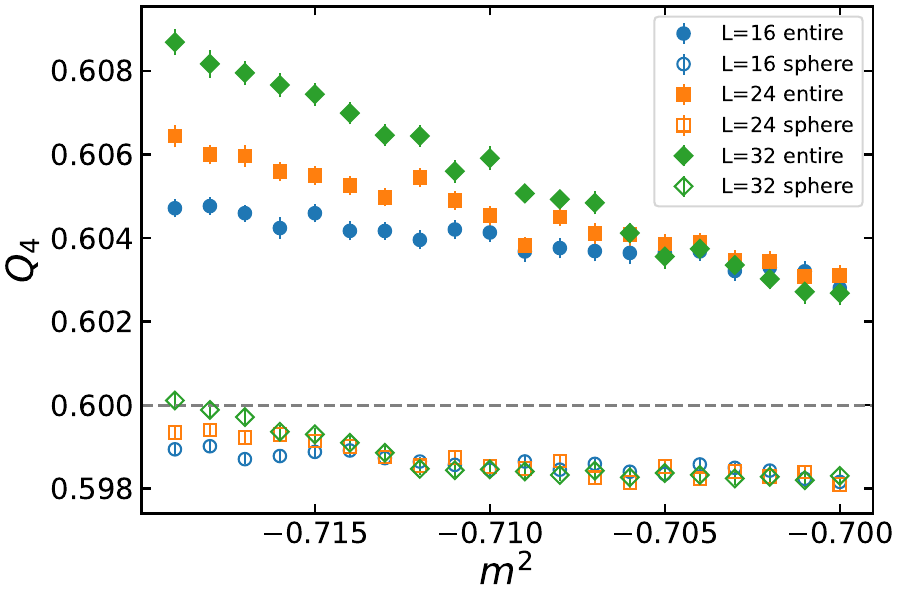}
    \includegraphics[width=0.45\linewidth, 
    alt={Graph depicting the anisotropy parameter in the immediate vicinity of the critical point, which grows slowly with system size.}
    ]{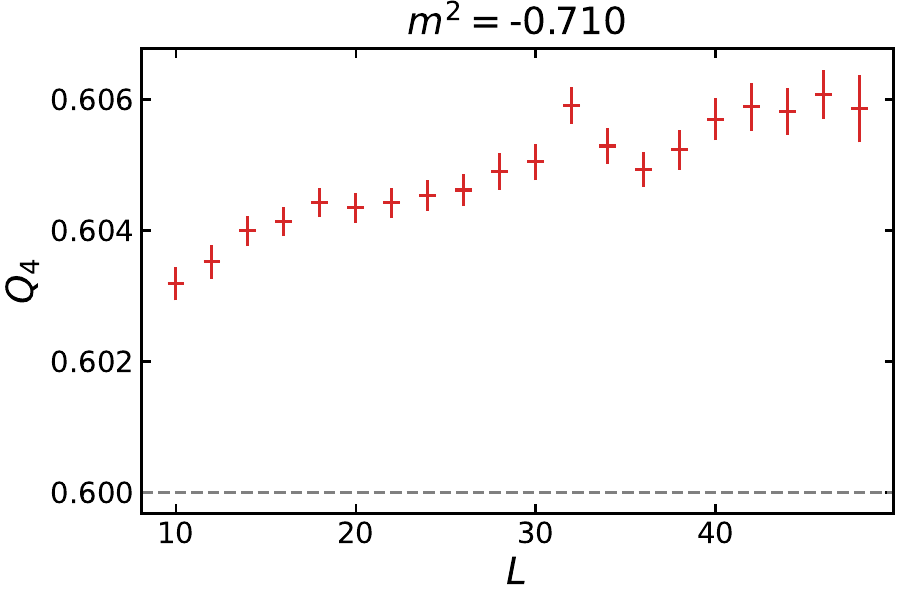}
    \caption{\label{fig_Q4_vs_m2_h000} 
        (Left) The anisotropy parameter $Q_4$ is plotted against $m^2$ near the critical point.
        The filled and open symbols depict the results of the entire lattice and spherical subregion, respectively. (The spherical subregion result was obtained by averaging over 8 non-overlapping spheres of radius $R=L/4$.)
        The blue-circle, orange-square, and green-diamond symbols correspond to the lattice sizes, $L=16$, $24$, and $32$, respectively.
        (Right) The entire lattice result of $Q_4$ at $m^2=-0.710$ is plotted against $L$.
        The gray broken line shows the isotropic limit $Q_4 = 0.6$.}
\end{figure}

To quantify this behavior, we compute the anisotropy parameter $Q_4$, Eq.~\eqref{eq:Q4}. At the critical point, $Q_4$ is expected to be a universal, scale-invariant quantity. However, we must distinguish between two distinct sources of rotational symmetry breaking: the lattice discretization itself and the toroidal geometry of the manifold we are working on (cubic lattice with periodic boundary conditions). Even for an ideal continuum theory with $O(3)$ rotational symmetry at criticality, the toroidal geometry can induce an apparent scale-invariant anisotropy along the axes of the torus.

To probe these effects, we measured $Q_4$ not only for the entire lattice but also within a spherical subregion of radius $R < L/2$.
In Fig.~\ref{fig_Q4_vs_m2_h000}, the left panel shows the $m^2$ dependence of $Q_4$ for $L=16$, $24$, and $32$, comparing the results from the entire lattice with those from the spherical subregion.
The right panel shows the $L$ dependence of $Q_4$ for the entire lattice at $m^2=-0.710$, which is the closest point to the critical value in our data.

Our results indicate that as the lattice size $L$ increases, the anisotropy parameter $Q_4$ tends toward a constant or varies only very slowly. 
The deviation from the isotropic limit of $Q_4=0.6$ is smaller when
sampling within the spherical region than on the entire lattice. 
This is consistent with a significant part of the deviation being due to the toroidal geometry. 
However, the spherical subregion is itself a finite region inside the torus, and its use is not expected to remove the effect of the torus entirely. 
It then remains unknown how much of the deviation in the spherical region is still due to the torus, and whether the breaking of rotational symmetry by the lattice action survives at long distances.
It would be interesting to separate the effect of the torus from a genuine
bulk breaking of rotational symmetry by performing the computation on a
geometry which is itself rotationally symmetric.

\section{Cubic anisotropy for \texorpdfstring{$h\neq 0$}{nonzero h} \label{sec:q4h}}

In Section \ref{sec:q4}, we observed that the breaking of the $O(3)$ symmetry at the critical point of model \eqref{eq:ourmodel} is very small and is due at least partly to the boundary conditions imposed by the cubic lattice.
This occurred despite the presence of the microscopic cubic deformations inherent to that model.
This observation naturally links to the question of whether the $O(3)$-symmetric dipolar fixed point is stable or unstable under explicit cubic deformations.
This question is also of experimental interest because the lattice structure might affect the nature of the phase transition in real materials.

There are two possibilities. Either all cubic deformations of the $O(3)$-symmetric dipolar fixed point are irrelevant, naturally explaining the emergence of $O(3)$ invariance despite the microscopic breaking. Or there are relevant deformations. In the latter case, the emergence of $O(3)$ invariance would be of an accidental nature, implying that the model \eqref{eq:ourmodel} is somehow fine-tuned.

While we do not attempt to map the entire phase diagram or definitively resolve the absolute stability of the fixed point here, we will attempt to shed some light on this question by introducing a further explicit cubic symmetry-breaking deformation \eqref{eq:hdef} to the potential, parameterized by coupling $h$, as already mentioned in Section \ref{sec:Model}. By making this coupling of the same order as the coupling $g$, we may hope to enhance the microscopic breaking to cubic symmetry. We will then look at the behavior of the $O(3)$ symmetry emergence indicators at the fixed point.

To do so, we performed Monte Carlo simulations of the model described in Section \ref{sec:Model} with the explicit $O(3)$ breaking term included. Each simulation consisted of $10^7$ sweeps (following $10^4$ thermalization sweeps), with measurements taken every $100$ sweeps.

\subsection{Simulation results \label{sec:hneq0sim}}

\begin{figure}[t]
    \centering
    \includegraphics[width=0.45\linewidth, 
    alt={Graph depicting the Binder ratio near the critical point for the negative symmetry-breaking parameter.}
    ]{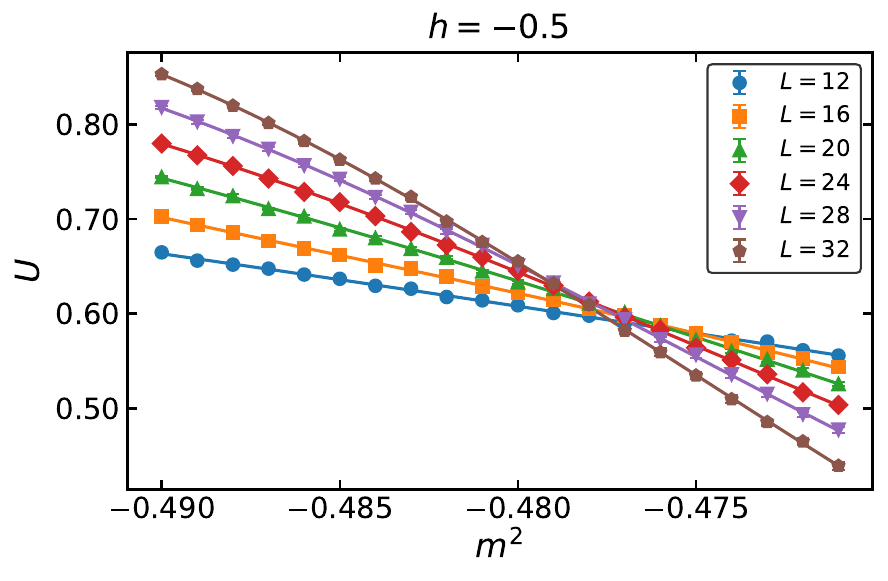}
    \includegraphics[width=0.45\linewidth, 
    alt={Graph depicting the Binder ratio near the critical point for the positive symmetry-breaking parameter.}
    ]{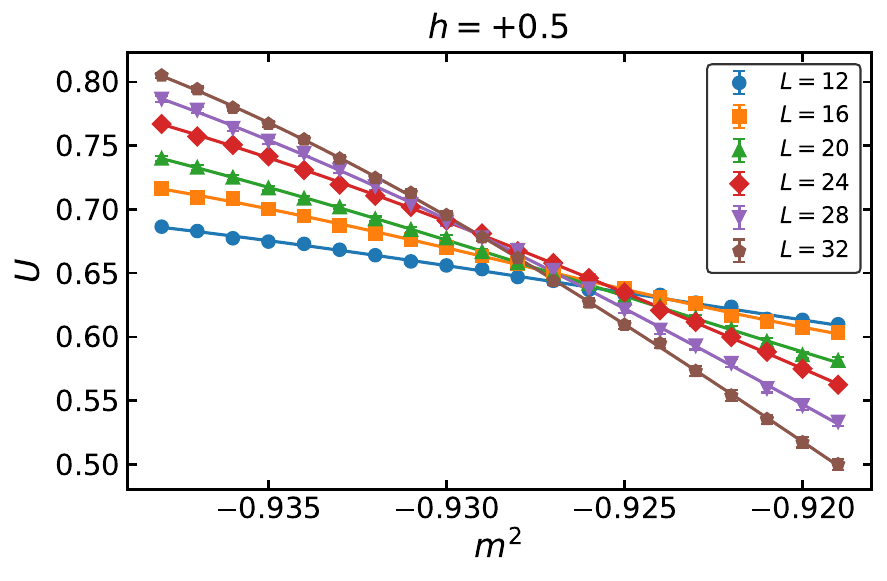}\\
    \includegraphics[width=0.45\linewidth, 
    alt={Graph showing the data-collapse result of the Binder ratio for the negative symmetry-breaking parameter.}
    ]{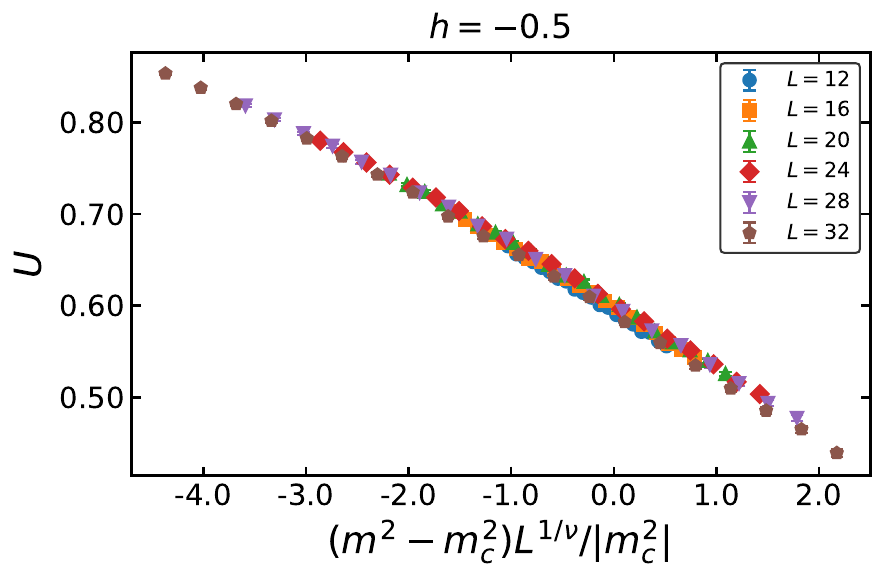}
    \includegraphics[width=0.45\linewidth, 
    alt={Graph showing the data-collapse result of the Binder ratio for the positive symmetry-breaking parameter.}
    ]{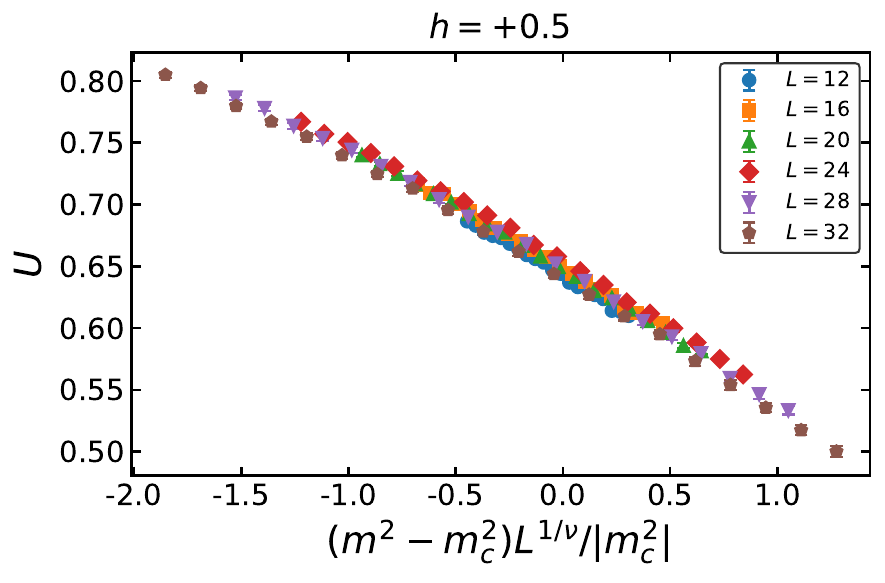}
    \caption{\label{fig:data_collapse_hne0} 
        Top: Binder ratio for $h=-0.5$ (left) and $h=+0.5$ (right).
        Bottom: Data collapse analysis of the Binder ratio for $h=-0.5$ (left) and $h=+0.5$ (right).}
\end{figure}

We performed Monte Carlo simulations at two distinct values of the $O(3)$ symmetry-breaking parameter, $h = \pm 0.5$, while fixing $g=1.0$. We first measured the Binder ratio$U$ for lattice sizes $L=12, 16, \dots, 32$. The results and the corresponding scaling collapse analysis are presented in Fig.~\ref{fig:data_collapse_hne0}. In the data collapse analysis shown in Fig.~\ref{fig:data_collapse_hne0}, the scaling ansatz for the Binder ratio is given by
\begin{equation}
    U = F\left[ L^{1/\nu}\frac{m^{2}-m_{c}^{2}}{|m_{c}|^{2}} \right],
    \label{eq_dc_ansatz_0}
\end{equation}
where $F$ is approximated by a cubic polynomial.  We perform the data collapse using the same methodology as in Section \ref{sec:Binder}.

Using the scaling ansatz without corrections in Eq.~\eqref{eq_dc_ansatz_0}, the fits yield $m_c^2 = -0.4773(1)$ and $\nu = 0.679(6)$ ($P=5.13$) for $h = -0.5$, and $m_c^2 = -0.9267(1)$ and $\nu = 0.690(5)$ ($P=6.30$) for $h = +0.5$.
We also tested ansatze A and B, which incorporate corrections to scaling. As shown in Table~\ref{tab_dc_U_h}, while the cost function $P$ formally improves, the statistical errors for the correction parameters become too large to yield reliable estimates.

\begin{table}[t]
    \centering
    \begin{tabular}{|c|c||c|c|c|c|c|}
        \hline
        $h$                     & ansatz                 & $m^2_c$       & $\nu$       & $\omega$   & $c_U$       & $P$ \tabularnewline
        \hline \hline
        \multirow{3}{*}{$-0.5$} & \eqref{eq_dc_ansatz_0} & $-0.4773(1)$  & $0.679(6)$  & $-$        & $-$         & $5.13$ \tabularnewline
        \cline{2-7}
                                & A                      & $-0.4803(4)$  & $0.695(11)$ & $0.13(17)$ & $-0.61(98)$ & $0.67$ \tabularnewline
        \cline{2-7}
                                & B                      & $-0.4805(4)$  & $0.744(10)$ & $0.05(13)$ & $-0.75(57)$ & $0.64$ \tabularnewline
        \hline \hline
        \multirow{3}{*}{$+0.5$} & \eqref{eq_dc_ansatz_0} & $-0.9267(1)$  & $0.690(5)$  & $-$        & $-$         & $6.30$ \tabularnewline
        \cline{2-7}
                                & A                      & $-0.9306(8)$  & $0.726(10)$ & $0.29(17)$ & $-0.40(15)$ & $1.58$ \tabularnewline
        \cline{2-7}
                                & B                      & $-0.9309(13)$ & $0.769(30)$ & $0.21(33)$ & $-0.49(32)$ & $1.53$ \tabularnewline
        \hline
    \end{tabular}
    \caption{\label{tab_dc_U_h}
        The optimized parameters obtained by the data collapse for $h = \pm 0.5$. The results using the scaling ansatz without corrections, Eq.~\eqref{eq_dc_ansatz_0}, are compared with those using ansatzes A and B.}
\end{table}

We should note that including these correction terms shifts the apparent critical point $m_c^2$ toward the ordered phase by a margin that exceeds the statistical errors.
This behavior is similar to the analysis at $h=0$, where the same scaling ansatz without corrections in Eq.~\eqref{eq_dc_ansatz_0} for lattice sizes $L=12, 16, \dots, 32$ yielded $m_c^2 = -0.70720(4)$ and $\nu = 0.691(3)$ with $P = 31.8$.

\begin{figure}[t]
    \centering
    \includegraphics[width=0.385\linewidth, 
    alt={Graph depicting the anisotropy parameters for the negative symmetry-breaking parameter, which are larger than the isotropic-limit value.}
    ]{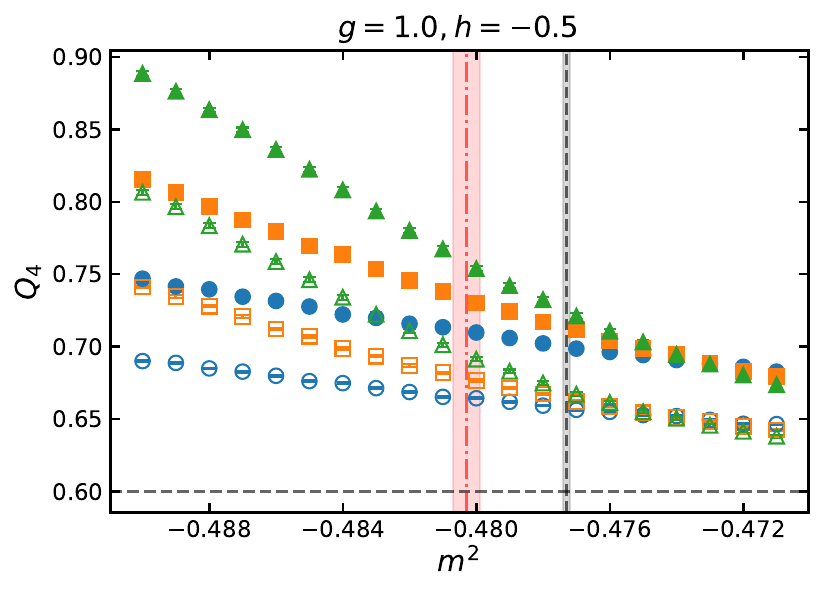}
    \includegraphics[width=0.515\linewidth, 
    alt={Graph depicting the anisotropy parameters for the positive symmetry-breaking parameter, which are smaller than the isotropic-limit value.}
    ]{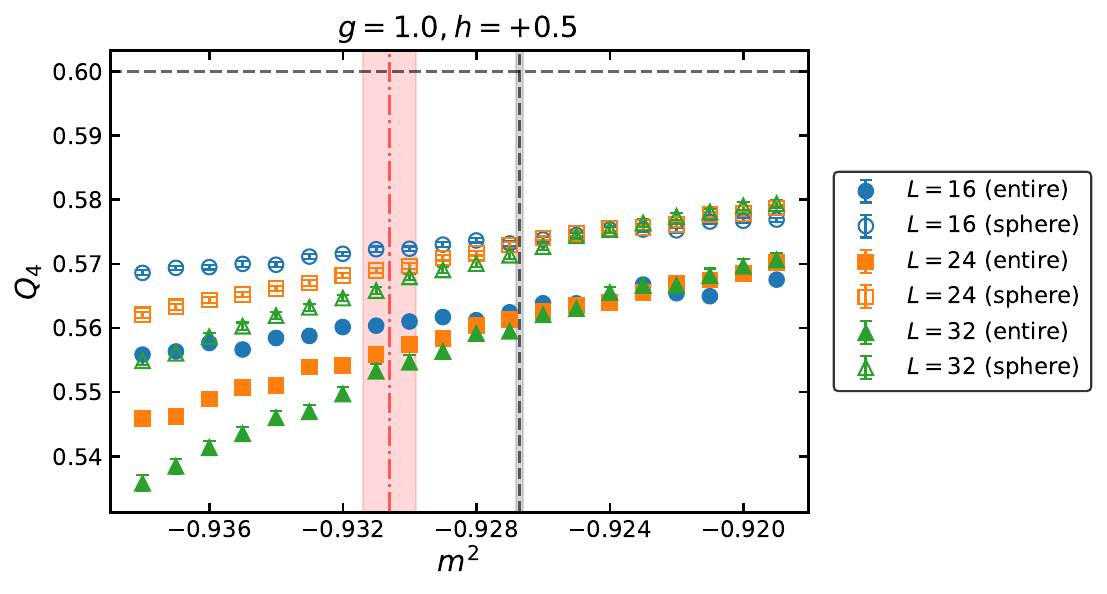}
    \caption{\label{fig:Q4_subvol} 
        The anisotropy parameter $Q_4$ for $h=-0.5$ (left) and $h=+0.5$ (right).
        The filled and open symbols depict the results for the entire lattice and the spherical subregion
        (same as in Fig.~\ref{fig_Q4_vs_m2_h000}), respectively.
        The blue circles, orange squares, and green diamonds correspond to lattice sizes $L = 16$, $24$, and $32$, respectively.
        The vertical black dashed lines indicate the apparent critical point $m_c^2$
        and its Jackknife error obtained from the fit without scaling corrections, Eq.~\eqref{eq_dc_ansatz_0}.
        The red dash-dotted lines represent the shifted critical point obtained using ansatz A, which incorporates scaling corrections.}
\end{figure}

The measurement of $Q_4$ in the presence of the additional cubic potential ($h = \pm 0.5$) is reported in Fig.~\ref{fig:Q4_subvol}.
Following the procedure in the previous section, we analyzed $Q_4$ not only over the entire lattice but also within a spherical subregion to probe boundary effects.

At the apparent critical point determined from the Binder ratio crossings, $Q_4$ clearly deviates from the isotropic limit ($Q_4=0.6$). The pattern of this deviation, $Q_4 > 0.6$ for negative $h$ and $Q_4 < 0.6$ for positive $h$, is consistent with mean-field theory expectations.
We further see that it exhibits a clear lattice-size dependence, unlike what we observed at $h=0$. For both $h = \pm 0.5$, the deviation of $Q_4$ from $0.6$ increases as $L$ increases. It should be noted that the systematic error in $m_c^2$ associated with the scaling corrections shifts the apparent critical point toward the ordered phase, where this $L$-dependence of $Q_4$ becomes even more pronounced. Since $Q_4$ at the critical point must be a scale-invariant quantity, this may indicate that the points $h = \pm 0.5$ are away from the fixed points even at $m^2 = m_c^2$ obtained from the Binder crossing. Note that as $Q_4$ increases with the volume, the system appears to evolve under the renormalization group flow in a direction that breaks $O(3)$ symmetry more and more. In the following subsection, we will discuss several RG interpretations of these findings.

\subsection{Interpretation \label{sec:interpretation}}

In this subsection, building upon the results presented in Sections \ref{sec:q4} and \ref{sec:hneq0sim}, we discuss possible RG flow diagrams for our lattice system in the presence of cubic anisotropy deformations.

Let us briefly review the known results from one-loop perturbation theory. Aharony and Fisher \cite{AharonyFisher} demonstrated that, within the perturbative $\epsilon$-expansion, the $O(3)$-symmetric dipolar fixed point is unstable under cubic anisotropy deformations in the potential, which corresponds to the naive continuum limit of our parameter $h$.
This was later confirmed at two loops by Bruce and Aharony \cite{PhysRevB.10.2078} who found the RG exponent (deviation from marginality) in $d=4-\epsilon$:
\beq
y_h=\Delta - d =  -\frac{3}{17}\epsilon-\frac{5978}{44217}\epsilon^2+\mathcal{O}(\epsilon^3)\,.
\label{eq:BA}
\eeq
These papers pointed out that there is a second anisotropy deformation appearing from the kinetic term, $\sum_i\partial_i \phi_i \partial_i \phi_i$, which is marginal at one-loop order. To our knowledge, its fate at higher loop orders has not been investigated.

Our main observation in Section \ref{sec:hneq0sim} is that for $h=\pm 0.5$, the deviation of the anisotropy parameter $Q_4$ from its $O(3)$-symmetric value is prominent. Furthermore, it exhibits a clear, monotonic $L$-dependence. In contrast, in Section \ref{sec:q4}, we found that for $h=0$, the $L$-dependence of $Q_4$ is much milder and may approach a constant in the large-$L$ limit. Moreover, this limiting value is very close to the $O(3)$-symmetric value. The slight deviation may be attributed to boundary conditions, though further efforts to minimize boundary effects are needed to draw definitive conclusions.
Note also that even when we set the bare coupling $h=0$, the underlying lattice discretization breaks the continuous $O(3)$ symmetry down to the discrete cubic group. Consequently, if the cubic perturbation is relevant, this inherent symmetry-breaking effect will grow under the RG flow.

\begin{figure}[t]
    \centering
    \includegraphics[width=0.3\linewidth, 
    alt={Figure showing the RG flow of the first scenario with the stable dipolar fixed point and two cubic fixed points.}
    ]{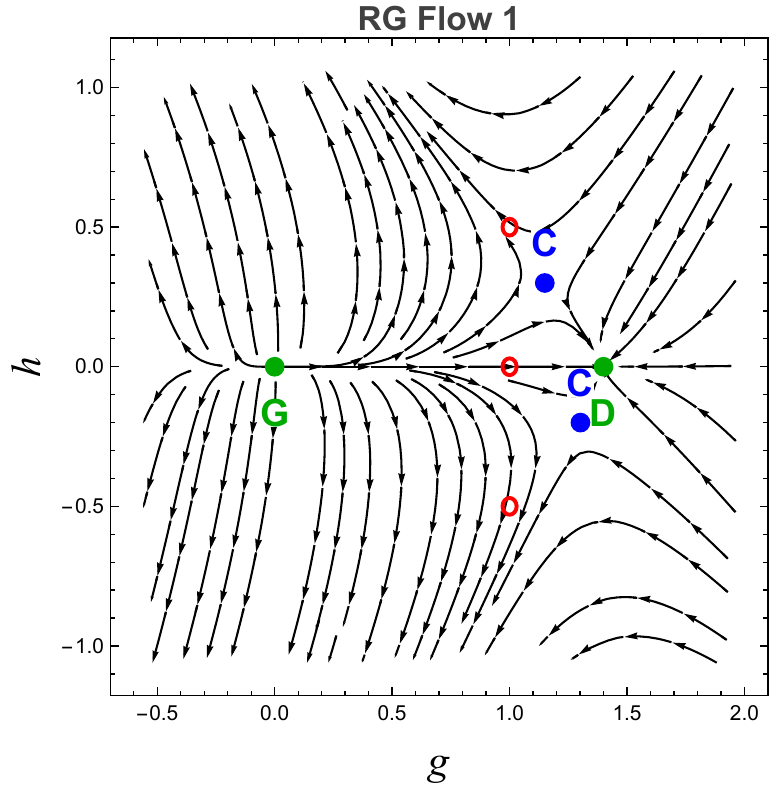}
    \includegraphics[width=0.3\linewidth, 
    alt={Figure showing the RG flow of the second scenario with the unstable dipolar fixed point and two cubic fixed points.}
    ]{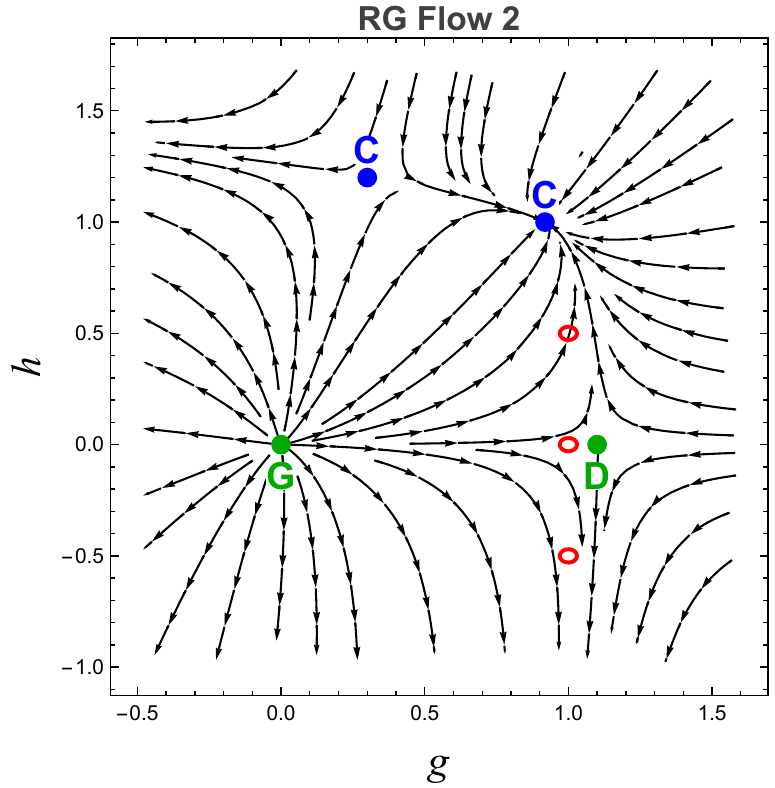}
    \includegraphics[width=0.3\linewidth, 
    alt={Figure showing the RG flow of the third scenario with the unstable dipolar fixed point but no cubic fixed point.}
    ]{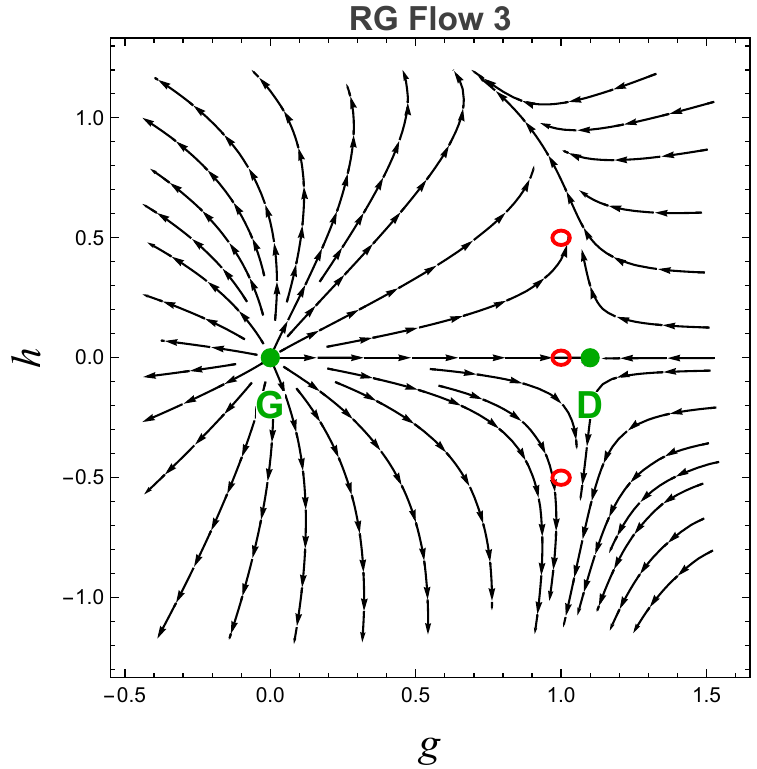}
    \caption{\label{fig:RG_flows} 
        Three possible RG flow diagrams in the presence of cubic anisotropy. 
        The left diagram depicts a scenario where the $O(3)$-symmetric dipolar fixed point is stable, 
        while the center diagram illustrates a scenario where it is unstable.
        In both diagrams, the green dots represent the $O(3)$-symmetric dipolar fixed point ($\mathsf{D}$) 
        and the Gaussian fixed point ($\mathsf{G}$). 
        The blue dots represent the additional cubic fixed points ($\mathsf{C}$). 
        The red circles indicate the hypothetical position of the three microscopic models ($h=0,\pm 0.5$) in this RG diagram. 
        (In this diagram, the $h=0$ line represents the line of enhanced symmetry. 
        Although the $h=0$ microscopic model is not exactly rotationally invariant, 
        our numerical simulations indicate that its deviation from rotational invariance is much smaller than that of the $h=\pm 0.5$ models, 
        and we place it on the horizontal line for illustration.) 
        The arrows indicate the direction of the RG flow in the coupling parameter space. 
        The right diagram illustrates a scenario where the $O(3)$-symmetric dipolar fixed point is unstable, and there are no cubic fixed points, 
        leading the RG flow to a first-order phase transition.}
\end{figure}

Prompted by these observations, we propose three possible scenarios for the RG flow diagram, which are illustrated in Fig.~\ref{fig:RG_flows}.
While these diagrams are illustrative rather than derived from non-perturbative calculations, they capture the expected RG flow patterns.
The two main assumptions are (1) the $m^2$ direction is strongly relevant and fine-tuned to be at the critical values; 
(2) we can model the RG flow as the flow in a two-dimensional space of couplings $(h,g)$.
All further couplings, such as e.g.~the aforementioned kinetic term cubic symmetry-breaking direction, are assumed irrelevant.

Finally, when preparing these toy model scenario plots, we limited ourselves to (3) beta-functions quadratic in the $g,h$ couplings.
Because of these, the number of fixed points is either two or four.
We also assumed (4) that the RG flow is a gradient flow.
These assumptions are not so essential; we made them for illustrative purposes only.

The three red circles in each of the figures represent the location of the three bare models, corresponding to $g=1.0$ with $h=0$ and $h = \pm 0.5$.
We must emphasize that setting the bare parameter $h=0$ on a finite lattice does not put the system exactly on the $O(3)$-symmetric invariant RG fixed line (represented by the $h=0$ line in the diagrams).
The lattice cutoff inevitably generates all operators compatible with cubic symmetry.
However, we assume this initial deviation is small and depict the $h=0$ bare model as if it were on the symmetric line.
We stress that this is an assumption made for the purpose of drawing the diagram.

The $h=0$ line in all the shown plots represents the submanifold of enhanced symmetry and hence is invariant under the RG flow.
The green dots on this line represent the position of two $O(3)$-symmetric fixed points - the trivial (Gaussian) and the nontrivial (dipolar).
Additional fixed points having only cubic symmetry are located at $h\ne 0$.
If present, these are shown as blue dots.
(In the third diagram, these two fixed points lie in the complex coupling plane and are therefore not shown.)

In the first scenario (Fig.~\ref{fig:RG_flows}, left), the $O(3)$-symmetric dipolar fixed point is stable against cubic perturbations.
However, the two additional cubic fixed points create a separatrix that bounds its basin of attraction.
If $h = \pm 0.5$ is sufficiently large, the bare parameters lie outside this basin, causing the RG flows to move away from the $O(3)$ fixed point.
This flow structure may explain the significant $L$-dependent deviations in $Q_4$ for $h = \pm 0.5$.
In this scenario, for $h=0$, the slight numerical increase of $Q_4$ with $L$ observed earlier should tend to a constant, which can be solely explained by boundary conditions.

This first scenario has tension with the perturbative result \eqref{eq:BA}, as it would require that $h$ turns around from relevant for small $\epsilon$ to be irrelevant for $\epsilon=1$.
One may wonder how likely this is.
One should keep in mind, though, that something similar does happen for the Heisenberg universality class, where the cubic deformation of the global $O(3)$ symmetry, while relevant at small $\epsilon$ \cite{aharony1973critical}, does become irrelevant at $\epsilon=1$ \cite{Chester:2020iyt, Hasenbusch:2022zur}, albeit only very weakly so.
So this turnaround could, in principle, happen for our model as well.

In the second scenario (Fig.~\ref{fig:RG_flows}, center), the $O(3)$-symmetric dipolar fixed point is unstable against cubic perturbations.
Flows for any $h \neq 0$ are repelled from the dipolar fixed point and driven toward the stable cubic fixed points or toward a first-order phase transition.
In this scenario, the $O(3)$ symmetry observed at $h=0$ would be due to an accidental fine-tuning.
In this scenario, for $h=0.5$ (or $h=-0.5$), if we increase the lattice size further, $Q_4$ should become constant, indicating scale-invariant behavior by reaching the stable fixed point.

In the third scenario (Fig.~\ref{fig:RG_flows}, right), the $O(3)$-symmetric dipolar fixed point is again unstable against cubic perturbations.
This scenario is a variant of the second scenario, but with no other fixed points present.
Thus, the $O(3)$ symmetry observed at $h=0$ would be due to a fine-tuning, and the phase transition at non-zero $h$ should be first-order rather than second-order for both signs of $h$.

In principle, we can also imagine a variant of the first scenario where the dipolar fixed point is stable, yet no cubic fixed points exist for finite values of the coupling $h$.
This was not compatible with the toy model assumptions (3), (4), but may well be possible for a more complicated beta-function dependence.
Additionally, it is possible that the cubic deformation in the kinetic term introduces another unstable direction; in such a case, the two-parameter diagrams here are insufficient, and reaching the $O(3)$-symmetric dipolar fixed point would require even further fine-tuning.

We leave the mapping of the entire phase diagram by studying the full $(h,g)$ plane (or with more relevant parameters if any) for future work.
Accurately resolving this global phase structure would also require reducing the effects of the boundary conditions and finite-size corrections.

\section{Discussion and outlook \label{sec:discussion-and-outlook}}

The main purpose of this work was to demonstrate that it is feasible to study the dipolar universality class via direct unbiased lattice Monte Carlo simulations.
For this, we introduced a model and a Markov Chain Monte Carlo algorithm that preserve the divergence-free constraint exactly.
Then we performed simulations up to moderately large volumes $L=48$ with modest computer resources and precisely determined the critical point of the phase transition.
While we have not aimed here at achieving high accuracy in critical exponents, we reported the first unbiased Monte Carlo determinations of the exponent $\nu$ and the critical value $F(0)$ of the Binder ratio.
We also studied whether rotational invariance emerges at the critical point, and the effect of the toroidal geometry on this question.
In addition, the model with the symmetry-breaking deformation was simulated, and we presented three conjectured RG flow diagrams.

In Table \ref{tab:critical_exponents}, we report our estimates of $\nu$ from Sections \ref{sec:Binder} (Binder fit) and \ref{sec:susceptibility} (Binder plus susceptibility) as well as the weak bound of $\eta$ from the latter fit.\footnote{
We recall the relation between the scaling dimensions of operators and the critical exponents $\Delta_{\phi_\mu} = (d-2)/{2} + \eta/2$, $\Delta_{\phi^2} = d -{1}/{\nu}$, where $d=3$.
The dimension of the least irrelevant singlet operator is related to $\omega$ via $\Delta_{\mathrm{irrel}} = d+\omega$.}
In the same table, we report prior determinations of the same critical exponents via RG, experiments, and Monte Carlo \cite{Itou:2025oku}.
As discussed in Section \ref{sec:susceptibility}, while $\eta$ cannot be determined due to the instability of the scaling fit, our determination of $\nu$ is likely to stand the test of time.
Our estimate of $\nu$ and bound of $\eta$ are compatible with the field theory and experimental results reported in Table \ref{tab:critical_exponents}.
In comparison, the previous Monte Carlo result \cite{Itou:2025oku} for $\nu$ showed a larger deviation.
It was likely due to the bias caused by the way of implementing the divergence-free constraint by an energy penalty, where the $\lambda \to \infty$ limit to ensure the exact constraint was not taken.
In the model of this paper, this bias is absent.

\begin{table}[t]
    \centering
    \renewcommand{\arraystretch}{1.2}
    \begin{tabular}{@{}llll@{}}
        \toprule
        \textbf{Method}                                    & $\nu$     & $\eta$       & $\gamma$  \\ %
        \midrule
        \multicolumn{4}{@{}l}{\textbf{RG}}                                                        \\
        3-loop \cite{2022NuPhB.98515990K}                  & 0.700(7)  & 0.033(8)     & 1.377(16) \\
        FRG (LPA, LPA') \cite{Nakayama:2023wrx}            & 0.752     & 0.04 (input) & 1.474     \\
        FRG (LPA') \cite{Kalagov_2026}                     & 0.7355    & 0.0423       & 1.440     \\

        \multicolumn{4}{@{}l}{\textbf{Experiment}}                                                \\
        EuO neutron scattering    \cite{PhysRevB.14.4908}  & 0.681(17) & 0.04 (input) & 1.387(36) \\
        EuS neutron scattering   \cite{PhysRevB.14.4908}   & 0.702(22) & 0.04 (input) & 1.399(40) \\
        EuO bulk magnetization    \cite{PhysRevB.3.1689}   & 0.675(5)  & 0.009(11)    & 1.29 (1)  \\
        EuO specific heat          \cite{PhysRevB.11.2678} & 0.681(3)  &              &           \\
        EuS specific heat         \cite{PhysRevB.17.282}   & 0.710(7)  &              &           \\

        \multicolumn{4}{@{}l}{\textbf{Monte Carlo}}                                               \\
        Energy penalty ($\lambda= 8$) \cite{Itou:2025oku}  & 0.601(8)  & 0.032(6)     & 1.183(16) \\
        This work (Binder)                                 & 0.735(27) &              &           \\
        This work (Binder + $\chi$)                        & 0.702(10) & $\ \lesssim 0.2$ &       \\
        \bottomrule
    \end{tabular}
    \caption{\label{tab:critical_exponents} 
        Dipolar universality class --- main critical exponents, from perturbative and non-perturbative RG, 
        from experiments, and from Monte Carlo. 
        Remarks: Experiment \cite{PhysRevB.14.4908} used $\eta=0.04$ as an input, but also reports the direct determination $\eta = 0.068$ for EuO. 
        Monte Carlo \cite{Itou:2025oku} implemented the divergence-free constraint via the energy penalty term $\lambda (\partial_\mu\phi_\mu)^2$.}
\end{table}

As is well known, the critical exponents of the dipolar model in 3D look uncannily close to those of the Heisenberg model, although their $4-\epsilon$ expansion is different.
See Table \ref{tab:critical_exponents-Heis} for the Heisenberg exponents from Monte Carlo and Conformal Bootstrap.
Hopefully, in the future, Monte Carlo simulations of the dipolar model can be pushed towards a similarly good level of accuracy.
One may start with our model and improve it by adding additional terms to eliminate leading corrections to scaling.
This strategy was used for the Ising and $O(N)$ universality classes by Hasenbusch \cite{Hasenbusch:2010hkh, Hasenbusch:2017yxr, Hasenbusch:2019gmx, Hasenbusch:2019jkj, Hasenbusch:2020pwj, Hasenbusch:2025yrl}; it should help here as well.
To tame the residual finite-size corrections, a simulation on the larger lattice size is certainly desirable, but we may have to develop a better updating algorithm, such as a cluster update compatible with the divergence-free constraint, to avoid the critical slow-down.

\begin{table}[t]
    \centering
    \renewcommand{\arraystretch}{1.2}
    \begin{tabular}{@{}llll@{}}
        \toprule
        \textbf{Method}                                & $\nu$       & $\eta$     & $\gamma$  \\
        \midrule
        Monte Carlo      \cite{Hasenbusch:2020pwj}     & 0.71164(10) & 0.03784(5) & 1.3964(2) \\
        Conformal bootstrap     \cite{Chester:2020iyt} & 0.71164(17) & 0.0379(1)  & 1.3963(4) \\
        \bottomrule
    \end{tabular}
    \caption{\label{tab:critical_exponents-Heis} 
        The Heisenberg universality class --- main critical exponents from Monte Carlo and conformal bootstrap. 
        We do not report RG and experiments, which have much larger error bars.}
\end{table}

The question of rotational invariance restoration also deserves further study.
In this paper, we could not cleanly eliminate the contribution from the shape of the manifold to this breaking.
To achieve this, one could repeat simulations in a spherical region with an open boundary condition.
Under such conditions, any measured breaking of rotational invariance would be intrinsically due to the bulk.

\section*{Acknowledgment}

We are grateful to Etsuko Itou and Atis Yosrakob for their collaboration at the early stages of this work.
We are grateful to Amnon Aharony for communications concerning the status of the cubic symmetry breaking of the $O(3)$-symmetric dipolar fixed point.
S.R.~thanks Martin Hasenbusch for discussions.
S.R.~is grateful to the Yukawa Institute of Theoretical Physics (Kyoto University) for the hospitality.
The numerical simulations were conducted using the PC clusters at RIKEN iTHEMS and at the Quantum Information Physics Group in Osaka Metropolitan University.
A.M.~is supported by JST-CREST No. JPMJCR24I1.
Y.N.~is in part supported by JSPS KAKENHI Grant Number 21K03581.
S.R.~is partially supported by the Simons Collaboration on the Probabilistic Paths to Quantum Field Theory (award SFI-MPS-PP-00012621-16).

\bibliography{reference.bib}
\bibliographystyle{ptephy}

\end{document}